\documentclass[a4paper,fleqn]{cas-dc}
\usepackage[square,numbers]{natbib}

\usepackage[switch,mathlines]{lineno}
\let\oldequation\equation
\let\oldendequation\endequation

\renewenvironment{equation}
  {\linenomathNonumbers\oldequation}
  {\oldendequation\endlinenomath}

\usepackage{xltabular}
\usepackage{multicol}

\usepackage{amssymb}

\ExplSyntaxOn
\RenewDocumentCommand \eadauthor {} 
    { 
      \seq_map_inline:Nn \l_stm_au_seq 
        { 
            \regex_extract_once:nnNTF {(\w)\w*-(\w)} { ##1 } \l_stm_au_fn_seq
            { 
                \seq_pop_left:NN \l_stm_au_fn_seq \temp_var
                \seq_use:Nn \l_stm_au_fn_seq { .- }
                { . } 
            }
            { 
                \regex_match:nnTF { \. } { ##1 } 
                { ##1 }
                { \tl_head:n {##1}. }
            }
      }{ ~\l_stm_au_sn_seq }
    }
\ExplSyntaxOff
\ExplSyntaxOn
\makeatletter

\RenewDocumentEnvironment { PrelimsAbstract } { O{} }
  {\parindent=0pt 
   { \fontsize{14pt}{16pt}\selectfont #1 }\par
   \vskip12pt
   { \fontsize{12pt}{14pt}\bfseries\selectfont\casprelimstitle } \par
   \vskip6pt
   \ifnum\theblind>0\relax
     \vspace*{\the\baselineskip}
   \else  
     \seq_use:Nn \g_stm_prelimsau_seq { ,\ }
   \fi  
   \vskip12pt
   \par 
  } 
  {}
  
\makeatother
\ExplSyntaxOff
\newif\ifabbreviation
\pretocmd{\thebibliography}{\abbreviationfalse}{}{}
\AtBeginDocument{\abbreviationtrue}
\DeclareRobustCommand\acroauthor[2]{%
  \ifabbreviation
    \ifcsname acroused@#2\endcsname
      #2%
    \else
      #1%
      ~[\mbox{#2}]% <----
      \expandafter\gdef\csname acroused@#2\endcsname{}%
    \fi
  \else
    \ifcsname bibacroused@#2\endcsname
        \mbox{#2}%
    \else
        \mbox{#1}~(\mbox{#2})%
        \expandafter\gdef\csname bibacroused@#2\endcsname{}%
    \fi
  \fi
}

\usepackage{lipsum}

\usepackage{tabularx}
\usepackage{booktabs}
\usepackage{longtable}
\usepackage{caption}
\usepackage{float}
\usepackage{ragged2e}
\usepackage{caption}

\newif\ifarxiv
\arxivtrue

\newif\iftodo
\usepackage{cleveref}
\crefname{subfigure}{\unskip}{\unskip}
\Crefname{subfigure}{\unskip}{\unskip}
\usepackage{subfigure}

\usepackage{siunitx}

\usepackage{xcolor}
\usepackage{hyperref}
\hypersetup{
    colorlinks=true,
    linkcolor=blue,
    filecolor=magenta,      
    urlcolor=cyan,
    pdftitle={Overleaf Example},
    pdfpagemode=FullScreen,
    }

\usepackage{tablefootnote}

\newif\ifshowchanges
\definecolor{lightgray}{RGB}{200,200,200}
\ifshowchanges
    \usepackage[commandnameprefix=always]{changes}
    \setcommentmarkup{\marginpar{\small\centering\color{purple}\textbf{[{#1}]}}}
    \setdeletedmarkup{ {\color{lightgray}\sout{#1}}}
\else
    \usepackage[final,commandnameprefix=always]{changes}
\fi

\DeclareSIUnit \decibelA {dB(A)}
\DeclareSIUnit \decibelC {dB(C)}
\DeclareSIUnit \soneGF {soneGF}
\DeclareSIUnit \acum {acum}
\DeclareSIUnit \asper {asper}
\DeclareSIUnit \vacil {vacil}
\DeclareSIUnit \tuhms {tuHMS}

\definecolor{set1_1}{RGB}{228,26,28}
\definecolor{set1_2}{RGB}{55,126,184}
\definecolor{set1_3}{RGB}{77,175,74}
\definecolor{den_1}{RGB}{239,209,0}
\definecolor{den_2}{RGB}{78,184,123}
\definecolor{den_3}{RGB}{0,127,196}
\definecolor{loaclr}{RGB}{152, 78, 163}
\definecolor{mclr}{RGB}{255, 127, 0}

\newcommand{\highlightlist}[1]{\vspace{-1.5\baselineskip}\textcolor{magenta}{#1}}

\newcommand{\dba}[1]{\SI{#1}{\decibelA}}
\newcommand{\dbc}[1]{\SI{#1}{\decibelC}}

\newcommand{\CheckedBox}[1]{%
  \ifnum#1=1
    \makebox[0pt][l]{\raisebox{0.15ex}{\hspace{0.2em}$\checkmark$}}%
  \fi
  $\square$%
}
\newcommand{\checkbox}[1]{\item[\CheckedBox{0}] #1}
\newcommand{\checkedbox}[1]{\item[\CheckedBox{1}] #1}

\newcommand{\Lx}[1]{$L_\text{#1}$}
\newcommand{\Lh}[2]{$L_\text{#1,\SI{#2}{\hour}}$}
\newcommand{\Lmin}[2]{$L_\text{#1,\SI{#2}{\minute}}$}
\newcommand{\Ls}[2]{$L_\text{#1,\SI{#2}{\second}}$}
\newcommand{\Lle}[2]{$L_\text{#1}\le\dba{#2}$}
\newcommand{\Lhle}[3]{$L_\text{#1,\SI{#2}{\hour}}\le\dba{#3}$}
\newcommand{\Lminle}[3]{$L_\text{#1,\SI{#2}{\min}}\le\dba{#3}$}

\newcommand{\dLhgeC}[4]{$\mu_{\Delta L_\text{#1,\SI{#2}{\hour}}} \in [{#3},{#4}] \;\dbc{}$}

\newcommand{\binL}{$m_\text{L}$}
\newcommand{\binR}{$m_\text{R}$}
\newcommand{\GRASIn}{$m_\text{in}$}
\newcommand{\GRASOut}{$m_\text{out}$}
\newcommand{\HDbinL}{$m_\text{L}^\text{HD}$}
\newcommand{\HDbinR}{$m_\text{R}^\text{HD}$}
\newcommand{\HDGRAS}{$m_\text{out}^\text{HD}$}
\newcommand{\NICUbinL}{$m_\text{L}^\text{NICU}$}
\newcommand{\NICUbinR}{$m_\text{R}^\text{NICU}$}
\newcommand{\NICUGRASOut}{$m_\text{out}^\text{NICU}$}
\newcommand{\NICUGRASIn}{$m_\text{in}^\text{NICU}$}

\newcommand{\mudiffLo}[4]{$\mu^{\Delta m_\text{#1,#2}}_{L_\text{#3}} = #4$}

\newcommand{\lmeart}{LME-ART-ANOVA}

\begin{document}
\let\WriteBookmarks\relax
\def\floatpagepagefraction{1}
\def\textpagefraction{.001}

\newcommand*{\papertitle}{Do neonates hear what we measure? Assessing neonatal ward soundscapes at the neonates' ears}
\shorttitle{\papertitle}    
\shortauthors{Lam et al.}  

\title[mode=title]{\papertitle}

\author[eee]{Bhan Lam}[orcid=0000-0001-5193-6560, degree=Ph.D.]
\ead{blam002@e.ntu.edu.sg}
\corref{c}\cortext[c]{Corresponding authors}
\credit{Conceptualization, Methodology, Software, Validation, Formal analysis, Investigation, Project administration, Data Curation, Writing - Original Draft, Writing - Review \& Editing, Visualization, Supervision}

\author[sgh]{Peijin Esther Monica Fan}[orcid=0000-0002-6325-154X, degree=RN]
%\ead{esther.monica.fan.p.j@sgh.com.sg}
\credit{Conceptualization, Methodology, Resources, Writing - Review \& Editing, Project administration}

\author[sgh]{Yih Yann Tay}[degree=RN]
%\ead{tay.yih.yann@sgh.com.sg}
\credit{Conceptualization, Investigation, Resources, Writing - Review \& Editing, Project administration}

\author[sghndm]{Woei Bing Poon}[degree=MRCPCH{,} FAMS]
%\ead{tay.yih.yann@sgh.com.sg}
\credit{Conceptualization, Investigation, Resources, Writing - Review \& Editing, Project administration}

\author[eee]{Zhen-Ting Ong}[orcid=0000-0002-1249-4760]
%\ead{ztong@ntu.edu.sg}
\credit{Resources, Investigation, Data Curation, Project administration}

\author[eee]{Kenneth Ooi}[orcid=0000-0001-5629-6275, degree=Ph.D.]
%\ead{wooi002@e.ntu.edu.sg}
\credit{Formal analysis, Resources, Writing - Review \& Editing}

\author[eee]{Woon-Seng Gan}[orcid=0000-0002-7143-1823, degree=Ph.D.]
%\ead{ewsgan@ntu.edu.sg}
\credit{Resources, Writing - Review \& Editing, Supervision, Supervision}

\author[sgh]{Shin Yuh Ang}[orcid=0000-0001-9614-7012,degree=MBA{,} RN]
\ead{ang.shin.yuh@singhealth.com.sg}
\corref{c}
%\cortext[c]{Corresponding authors}
\credit{Conceptualization, Resources, Investigation, Writing - Review \& Editing, Supervision}

\affiliation[eee]{
    organization={%%
        School of Electrical and Electronic Engineering, 
        Nanyang Technological University%
    },
    addressline={50 Nanyang Ave}, 
    postcode={639798}, 
    country={Singapore}
}

\affiliation[sgh]{
    organization={%%
        Nursing Division, 
        Singapore General Hospital%
    },
    addressline={Outram Rd, 169608}, 
    country={Singapore}
}

\affiliation[sghndm]{
    organization={%%
        Department of Neonatal and Developmental Medicine, 
        Singapore General Hospital%
    },
    addressline={Outram Rd, 169608}, 
    country={Singapore}
}

\begin{abstract}
Acoustic guidelines for neonatal intensive care units (NICUs) aim to protect vulnerable neonates from noise-induced physiological harm. However, the lack of recognised international standards for measuring neonatal soundscapes has led to inconsistencies in instrumentation and microphone placement in existing literature, raising concerns about the relevance and effectiveness of these guidelines. This study addresses these gaps through long-term acoustic measurements in an operational NICU and a high-dependency ward. We investigate the influence of microphone positioning, bed placement, and ward layout on the assessment of NICU soundscapes. Beyond traditional A-weighted decibel metrics, this study evaluates C-weighted metrics for low-frequency noise, the occurrence of tonal sounds (e.g., alarms), and transient loud events known to disrupt neonates' sleep. Using linear mixed-effects models with aligned ranks transformation ANOVA (LME-ART-ANOVA), our results reveal significant differences in measured noise levels based on microphone placement, highlighting the importance of capturing sound as perceived directly at the neonate's ears. Additionally, bed position and ward layout significantly impact noise exposure, with a NICU bed position consistently exhibiting the highest sound levels across all (psycho)acoustic metrics. These findings support the adoption of binaural measurements along with the integration of additional (psycho)acoustic metrics, such as tonality and transient event occurrence rates, to reliably characterise the neonatal auditory experience.
\end{abstract}

% \includegraphics[width=\linewidth]{figures/grabs_placeholder.pdf} 

\ifarxiv\else
\begin{highlights}
\item Binaural microphone positioning reveals disparities in neonatal noise exposure
\item Ward layout and bed positioning significantly impacts neonatal soundscapes
\item Frequent loud transients despite lower averages highlight need for advanced metrics
\item C-weighted metric differences suggest the need for low-frequency monitoring
\end{highlights}
\fi

\begin{keywords}
neonatal intensive care \sep indoor noise \sep hospital soundscape \sep binaural \sep hospital acoustics \sep building acoustics

\end{keywords}

\maketitle

\ifarxiv\else\linenumbers\fi

\section{Introduction}
\label{sec:Introduction}

\subsection{Background and Motivation}

\iftodo
\begin{itemize}[leftmargin=*]
  \checkedbox{neonate hearing biology}
  \begin{itemize}
      \checkbox{frequency range}
  \end{itemize}
  \checkedbox{impact of noise on neonates}
\end{itemize}
\fi

The development of the auditory system begins very early in gestation. By 26 weeks of gestation, the human fetus begins to react to auditory stimuli. The auditory system then continues to develop to be fine-tuned to specific frequencies, to process acoustic stimuli into electric signals sent to the brainstem, to discern different speech phonemes and to facilitate learning and memory formation \cite{mcmahon_auditory_2012}. \chadded[comment=R3.1]{Following birth, neonates become increasingly attuned to their acoustic surroundings. They can distinguish auditory cues from background noise and differentiate human speech from non-biological sounds as early as 28 weeks \cite{kuhn_auditory_2017}.}
 
While sensory input due to the acoustic environment has an impact on the structure and function of the neural systems throughout life, its influence is most pronounced during infancy \cite{Dahmen2007}. Hence, the auditory environment of a developing infant is extremely important to its development. 

In the uterus, the fetus receives auditory input via bone conduction and is surrounded by fluid and maternal tissues which attenuates high frequency sounds more so than low frequency sounds \cite{Lahav2015}. Preterm exit from the optimal environment of the womb means that preterm infants are exposed to hearing via air conduction, as well as sounds of higher frequencies and higher volumes (especially in the Neonatal Intensive Care Unit (NICU)). \chreplaced[]{Prolonged exposure to high-frequency noise, especially in NICUs, can disrupt physiological and developmental processes. Studies have shown that such noise may lead to alterations in blood pressure, respiratory rates, and sleep cycles, all of which could be detrimental to neural growth and auditory development \cite{Lahav2015, lahav_acoustic_2014, el-metwally_potential_2020, retsa_longstanding_2024}.}{This may be harmful to infants as high frequency noise exposure has shown to alter blood pressure, decrease respiratory rate and disrupt sleep \cite{Lahav2015}.}

\chadded[]{In the NICU, loud transient sounds can induce immediate physiological responses, including elevated heart rates, altered respiratory patterns, and disrupted sleep cycles \cite{wachman_effects_2011, kuhn_infants_2012, shimizu_sound_2016}. These sudden noises are likely to trigger startle reflexes and stress-mediated defensive response, which may interfere with homeostatic regulation and delay neurological development \cite{kuhn_infants_2012, Philbin2017}.}

\chadded[comment=R3.1 R4.2]{The absence of \textit{in utero} maternal sounds increases the risk of long-term speech and language deficits in preterm neonates \cite{mcmahon_auditory_2012, webb_mothers_2015}.  Evidence suggests that interventions involving the introduction of spoken or recorded maternal sounds, can mitigate these effects, promoting improved auditory and visual orientation as well as enhanced cognitive and language outcomes later in life \cite{philpott-robinson_impact_2017}. Consequently, optimizing the NICU soundscape by minimizing non-biological noise is essential, as it may enhance the efficacy of maternal voice interventions \cite{kuhn_auditory_2017,el-metwally_potential_2020}. However, a recent systematic review found that no NICU meets the American Academy of Pediatrics guidelines of \dba{45}, indicating a need for improved NICU designs to support auditory and developmental health \cite{andy_systematic_2025}.}

\subsection{Acoustic design guidelines and regulations in NICUs}

\iftodo
\begin{itemize}[leftmargin=*]
  \checkedbox{WHO and AAP}
  \checkedbox{NICU design guidelines 9th ed}
  \checkedbox{Singapore guidelines}
  \checkedbox{Summary table}
  \checkedbox{incubator maximum sound level and standard}
\end{itemize}
\fi

In the context of neonatal ward soundscapes, it is crucial to adhere to established and updated guidelines and regulations to ensure an optimal acoustic environment to protect the well-being of vulnerable neonates. Among the various guidelines available, the World Health Organisation (WHO) followed by American Academy of Pediatrics (AAP) guidelines are frequently referenced in the literature \citep{DeLimaAndrade2021}. The WHO recommends stringent acoustic standards for hospital wards, including an A-weighted equivalent sound pressure level (SPL) indoors of \Lhle{AF}{24}{30} and an instantaneous maximum SPL level of \Lle{AFmax}{40} \citep{Berglund1999a}, where the subscript $\{{\cdot}\}_\text{F}$ refers to the `\textit{fast}' time response (\SI{125}{\milli\second}) according to IEC~61672\nobreakdash‑1 \citep{IEC6167212013}. Similarly, the AAP Guidelines suggests maintaining sound pressure levels (SPL) below a day-night averaged SPL (\dba{10} penalty to nighttime levels) of \Lle{dn}{45} \citep{AAP1997}, based on the U.S. Environmental Protection Agency guidelines for hospitals \citep{EPA1974}. 

While acknowledging the widespread referencing of the WHO (\citeyear{Berglund1999a}) and AAP (\citeyear{AAP1997}) guidelines in the literature, it is important to recognise the possibility that these guidelines may now be outdated. Consequently, in response to the need for more intuitive and realistic standards, the Consensus Committee (CC) on \textit{Recommended Design Standards for Advanced Neonatal Care} has consistently revised the ``\textit{Recommended Standards for Newborn ICU Design}''. The most recent update, the 10\textsuperscript{th} edition in 2023 \citep{altimier_recommended_2023}, reflects these changes aimed at enhancing the standard's practicality and alignment with current research and practices in neonatal care.

According to the CC, the recommended one-hour SPLs are \Lhle{AS50}{1}{50} and \Lhle{AS10}{1}{65}, as measured three feet from an infant bed or other relevant listening position. Here, subscript $\{{\cdot}\}_\text{S}$ refers to the `\textit{slow}' (\SI{1}{\second}) time response, while $\{{\cdot}\}_{50}$ and $\{{\cdot}\}_{10}$ represent the percentage exceedance levels as described in ISO~1966\nobreakdash-1 \cite{ISO1996-1}. Notably, the $9^\text{th}$ edition update replaced the equivalent SPL guidelines of \Lhle{AS}{1}{45} to the \SI{50}{\percent} exceedance levels of \Lhle{AS50}{1}{50}; increased the $L_\text{AS10,\SI{1}{\hour}}$ from \dba{50} to \dba{65}; and removed the maximum SPL requirement of \Lle{ASmax}{65}. Various portions of the CC NICU design standards have been adopted in American Institute of Architects/Facilities Guidelines Institute Guidelines \citep{FGI2022}, AAP/American College of Obstetricians and Gynecologists’ (ACOG) Guidelines for Perinatal Care \citep{AAPACOG2017}, and forms the basis for noise emission limits in infant incubators in IEC 60601-2-19 \citep{IEC60601-2-19}.

In Singapore, there are no specific guidelines or regulations governing the acoustic environment in NICUs or buildings in general. However, under the Environmental Protection and Management Act of 1999, two statutory regulations control noise levels emitted from construction sites and factory premises, with specific limits for noise sensitive buildings such as hospitals. For construction noise \citep{AGCConst2008}, the maximum permissible levels for hospitals as measured \SI{1}{\meter} from the facade, are as follows: (1) \Lhle{AF}{12}{60} between 7am to 7pm, (2) \Lhle{AF}{12}{50} between 7pm to 7am, (3) \Lminle{AF}{5}{75} between 7am to 7pm, and (4) \Lminle{AF}{5}{55} between 7pm to 7am. Regarding factory premises, the maximum permissible levels measured on the boundary facing the hospital are: (1) \Lhle{AF}{12}{65} between 7 am and 7 pm, (2) \Lhle{AF}{3}{55} between 7 pm and 11 pm, and (3) \Lhle{AF}{9}{50} between 11 pm and 7 am \citep{AGCFactoryNoise2008}. 

Another point of reference for NICU acoustic design requirements in Singapore can be found in the BCA Green Mark 2021 (GM: 2021) green building certification for healthcare facilities. The acoustic requirements in GM:2021 are derived from guidelines provided by the United Kingdom (UK) \citep{HTM0801_2013} and a joint standard established by Australia and New Zealand (ANZ) \citep{ASNZS2107_2016}. The Healthcare Technical Memorandum (HTM~08\nobreakdash-01) from the UK, sets limits for external sound sources within multi-bed wards, with recommended levels of \Lhle{AF}{1}{40} during the day and \Lhle{AF}{1}{35} at night (11pm to 7am), as well as \Lle{AFmax}{45}. Noise rating (NR) limits were also specified for mechanical and electrical services within the ward (i.e. $\text{NR}\le30$). In contrast to the broader guidelines applicable to hospital wards in general, the ANZ standard (AS/NZS~2107) provides specific interior noise limits tailored for NICUs and pediatric ICUs (PICU). These standards define acoustic targets such as \Lle{ASmax}{50} for external transient noise, as well as interior ambient noise levels of \Lhle{AS}{1}{45} and \Lhle{AS10}{1}{50} for NICU/PICU wards. The recommended acoustic limits and guidelines discussed are summarised in \Cref{tab:sound-standards}.

%\footnotesize
%\tiny
\begin{table}[ht]
\footnotesize
  \centering
  \caption{Acoustic standards and guidelines for Neonatal Intensive Care Units (NICU) / Pediatric Intensive Care Units (PICU)}
  \label{tab:sound-standards}
  \begin{tabularx}{\linewidth}{p{1cm}X*{2}{>{\raggedright\arraybackslash}p{2.55cm}}}
    \toprule
    \textbf{Source}\textsuperscript{\textit{1}}
    & \textbf{NICU/\newline PICU} 
    & \textbf{Ambient Sound Level} 
    & \textbf{External Sound Intrusion} \\
    \midrule
    WHO \citep{Berglund1999a}
    & No 
    & \Lhle{AF}{24}{30}\newline \Lle{AFmax}{40}
    & - \\
    \midrule
    AAP \citep{AAP1997}
    & Yes 
    & \Lle{dn}{45}
    & - \\
    \midrule
    CC (8\textsuperscript{th} ed) \citep{White2013}
    & Yes
    & \Lhle{AS}{1}{45}\newline \Lhle{AS10}{1}{50}\newline \Lle{ASmax}{65}
    & - \\
    \midrule
    CC (9 \& 10\textsuperscript{th} ed) \citep{White2020,altimier_recommended_2023}
    & Yes
    & \Lhle{AS50}{1}{50}\newline \Lhle{AS10}{1}{65}
    & - \\
    \midrule
    HTM 08-01 \citep{HTM0801_2013}
    & No 
    & $\text{NR}\le30$ 
    & \Lhle{AF}{1}{40}\newline (7am to 11pm)\newline \Lhle{AF}{1}{35}\newline (11pm to 7am)\newline \Lle{AFmax}{45}\\
    \midrule
    AS/NZS 2107 \citep{ASNZS2107_2016}
    & Yes 
    & \Lhle{AS}{1}{45}\newline\Lhle{AS10}{1}{50}
    & \Lle{ASmax}{50}\\
    \midrule
    IEC 60601-2-19 \citep{IEC60601-2-19}
    & Yes
    & \Lle{ASmax}{60} (ambient sound inside incubator)\newline \Lle{ASmax}{80} (alarm sounding inside incubator)
    & -\\
    \bottomrule
  \end{tabularx}
  \noindent
  \begin{minipage}{1\linewidth}
    \scriptsize
    \textsuperscript{\textit{1}}World Health Organisation (WHO), American Academy of Pediatrics (AAP), Consensus Committee (CC) on recommended design standards for advanced neonatal care, HTM (Health Technical Memoranda), Australian/New Zealand Standard (AS/NZS), International Electrotechnical Commission (IEC) 
  \end{minipage}
\end{table}

\iftodo
\subsection{Neonatal intensive care unit (NICU) soundscape}
\begin{itemize}[leftmargin=*]
  \checkedbox{state of noise in neonatal wards}
  \checkedbox{incubator acoustics}
\end{itemize}
\fi

Despite notable advancements in noise control technology, a growing body of literature highlights the increasing noise levels within hospitals \cite{DeLimaAndrade2021,Busch-Vishniac2019, Busch-Vishniac2023,Lam2022c}. Within neonatal intensive care units (NICUs), the role of acoustics becomes even more critical as it directly impacts the survival rates and recovery process of vulnerable neonates. Previous measurements of A-weighted equivalent sound pressure levels (SPL) near the ears of infants in open-box incubators have revealed values ranging from \num{54.7} to \SI{60.44}{\decibelA}, surpassing the recommended guidelines outlined in \Cref{tab:sound-standards}. 

 Furthermore, neonates are frequently exposed to loud transient sounds, reaching levels between \num{62}--\SI{90.6}{\decibelA}, as indicated by variations in the \Lx{Amax} indicators presented in \Cref{tab:niculevels}. These transient noises can be attributed to various sources, including medical-staff activities, medical devices, or alarm peaks \citep{Bertsch2020}. Such high sound levels pose potential risks to the delicate hearing system of neonates, with \Lx{Amax} levels reaching up to \SI{94.8}{\decibelA} inside incubators \citep{Parra2017}, potentially resulting in permanent cochlear damage \citep{Surenthiran2003} or temporary hearing threshold shifts \citep{McCullagh1979}. Moreover, sudden sounds can trigger startle responses in neonates, interrupting their rest and recovery \citep{Philbin1999}. It is worth noting that the \Lh{AS10}{1} measurements reported by \citeauthor{Krueger2007} fall well within the NICU design guidelines of the 9\textsuperscript{th} edition (\Lhle{AS10}{1}{65}) ranging from 59.26--\SI{60.6}{\decibelA}.

\begin{table*}[ht]
  \centering
  \caption{A summary of acoustic measurement studies in neonatal intensive care units (NICU) with reported acoustic indicators and measurement equipment standards.}
  \label{tab:niculevels}
  \begin{tabularx}{\textwidth}{%
  >{\raggedright\arraybackslash}p{0.1\textwidth}%
  >{\raggedright\arraybackslash}p{0.1\textwidth}%
  >{\raggedright\arraybackslash}p{0.1\textwidth}%
  >{\raggedright\arraybackslash}p{0.12\textwidth}%
  >{\raggedright\arraybackslash}p{0.28\textwidth}X}
    \toprule
    \textbf{Source} & \textbf{Indicator} & \textbf{Value} [\si{\decibelA}] & \textbf{Measurement Period} & \textbf{Measurement Equipment Placement} & \textbf{Measurement Equipment} \\
    \midrule
    \multirow[t]{6}{0.1\textwidth}{\citeauthor{Krueger2007} \citeyear{Krueger2007}} 
    & \Lh{AS}{1} 
    & 60.44 
    & \multirow[t]{3}{0.12\textwidth}{Before: \SI{8}{\hour}/day for 9 days}
    & \multirow[t]{6}{0.28\textwidth}{Behind bed spaces 2--3 feet above countertop}
    & \multirow[t]{6}{*}{SLM (Class 1)} 
    \\
    \cline{2-3}
    & \Lh{AS10}{1} 
    & 59.26 &  &  & 
    \\
    \cline{2-3}
    & \Lh{ASmax}{1} & 78.39 &  &  & 
    \\
    \cline{2-4}
    & \Lh{AS}{1} 
    & 56.4 
    & \multirow[t]{3}{0.12\textwidth}{After: \SI{8}{\hour}/day for 2 days}
    &  
    & 
    \\
    \cline{2-3}
    & \Lh{AS10}{1} & 60.6 &  &  & 
    \\
    \cline{2-3}
    & \Lh{ASmax}{1} & 90.6 &  &  & 
    \\
    \midrule
    \citeauthor{Darcy2008} \citeyear{Darcy2008}
    & \Ls{AS}{1}
    & 53.9--60.6
    & 5/\si{\hour} in \SI{2}{\hour} 
    & Center of bay 
    & SLM (Class 2) 
    \\
    \midrule
    \multirow[t]{2}{0.1\textwidth}{\citeauthor{Lahav2015} \citeyear{Lahav2015}} 
    & \Lh{A}{24} 
    & 60.05
    & \multirow[t]{2}{*}{5 days} 
    & \multirow[t]{2}{*}{Center of bay}
    & \multirow[t]{2}{*}{SLM (Class 1)} 
    \\
    \cline{2-3}
    & \Lh{A}{24} 
    & 58.67 
    & 
    & 
    &  
    \\
    \midrule
    \multirow[t]{2}{0.1\textwidth}{\citeauthor{Romeu2016} \citeyear{Romeu2016}} 
    & \Ls{A}{1} 
    & 53.4--62.5
    & \SI{56}{\hour}
    & Inside incubator (roof)
    & \multirow[t]{2}{*}{SLM (Class 1)} 
    \\
    \cline{2-5}
    & \Ls{A}{1}
    & 60.4--65.5 
    & \SI{56}{\hour}
    & Outside incubator
    &  
    \\
    \midrule
    \multirow[t]{3}{0.1\textwidth}{\citeauthor{Shoemark2016} \citeyear{Shoemark2016}} 
    & \Lx{AS} 
    & 58 
    & \multirow[t]{3}{*}{720/\si{\hour} in \SI{24}{\hour}} 
    & \multirow[t]{3}{*}{Near head-end of bed} 
    & \multirow[t]{3}{*}{SLM (Class 1)} 
    \\
    \cline{2-3}
    & \Lx{ASmax} & 62 &  &  &  
    \\
    \cline{2-3}
    & \Lx{ASmin} & 55 &  &  &  
    \\
    \midrule
    \multirow[t]{2}{0.1\textwidth}{\citeauthor{Park2017neonatal} \citeyear{Park2017neonatal} } 
    & \Lmin{A}{10}
    & 56.7 
    & \SI{24}{\hour}
    & Inside incubator 
    & \multirow[t]{2}{*}{IEC 61094-4 WS2P} 
    \\
    \cline{2-5}
    & \Lmin{A}{10} 
    & 57.1
    & \SI{24}{\hour}
    & Outside incubator
    &  
    \\
    \hline
    \multirow[t]{6}{0.1\textwidth}{\citeauthor{Parra2017} \citeyear{Parra2017}} 
    & \Lx{A} 
    & 59.5 
    & \multirow[t]{3}{*}{3600/\si{\hour} in \SI{24}{\hour}} 
    & \multirow[t]{3}{*}{Center of bay}
    & \multirow[t]{6}{*}{Dosimeter (Class 2)} \\
    \cline{2-3}
    & \Lx{A10} & 61.8 &  &  &  \\
    \cline{2-3}
    & \Lx{Amax} & 85.2 &  &  &  \\
    \cline{2-5}
    & \Lx{A} 
    & 65.8 
    & \multirow[t]{3}{*}{3600/\si{\hour} in \SI{24}{\hour}}
    & \multirow[t]{3}{*}{Inside incubator, \SI{30}{\centi\meter} from the ears}
    &  
    \\
    \cline{2-3}
    & \Lx{A10} & 68.1 &  &  &  \\
    \cline{2-3}
    & \Lx{Amax} & 94.8 &  &  &  \\
    \midrule
    \multirow[t]{2}{0.1\textwidth}{\citeauthor{Smith2018} \citeyear{Smith2018}} 
    & \Ls{AF}{1} 
    & 58.1
    & 60/\si{\hour} in \SI{4}{\hour}
    & Open-pod; between two Isolette  
    & \multirow[t]{2}{*}{Dosimeter (Class 2)} \\
    \cline{2-5}
    & \Ls{AF}{1} 
    & 54.7 
    & 60/\si{\hour} in \SI{4}{\hour}
    & Private room; near ears 
    & \\
    \midrule
    \multirow[t]{2}{0.1\textwidth}{This study}
    & \Lx{AS,1h} & 56.09 (3.48) & \multirow[t]{4}{*}{\SI{981}{\hour}} & \multirow[t]{4}{0.28\textwidth}{Binaural microphone affixed to ears on neonate doll in a open bassinet (HD-A)} &  IEC 61094-4 WS2F \\
    \cline{2-3}
    & \Lx{AS10,1h} & 58.98 (3.87) &  &  &  \\
    \cline{2-3}
    & \Lx{AS50,1h} & 51.59 (3.36)  &  &  &  \\
    \cline{2-3}
    & \Lx{ASmax,1h} & 73.76 (4.20) &  &  &  \\
    \cline{2-5}
    & \Lx{AS,1h} & 59.38 (1.90) & \multirow[t]{4}{*}{\SI{392}{\hour}} & \multirow[t]{4}{0.28\textwidth}{Binaural microphone affixed to ears on neonate doll in a open-type incubator (NICU-A)} &  \\
    \cline{2-3}
    & \Lx{AS10,1h} & 61.40 (2.70) &  &  &  \\
    \cline{2-3}
    & \Lx{AS50,1h} & 57.63 (1.38)  &  &  &  \\
    \cline{2-3}
    & \Lx{ASmax,1h} & 73.4 (4.12) &  &  &  \\
    \cline{2-5}
    & \Lx{AS,1h} & 58.29 (1.81) & \multirow[t]{4}{*}{\SI{98}{\hour}} & \multirow[t]{4}{0.28\textwidth}{Binaural microphone affixed to ears on neonate doll in a open-type incubator (NICU-B)} &  \\
    \cline{2-3}
    & \Lx{AS10,1h} & 59.89 (2.27) &  &  &  \\
    \cline{2-3}
    & \Lx{AS50,1h} & 56.80 (1.15)  &  &  &  \\
    \cline{2-3}
    & \Lx{ASmax,1h} & 71.47 (4.81) &  &  &  \\
    \bottomrule
\end{tabularx}
\end{table*}

\subsection{NICU soundscape assessment}

\iftodo
\begin{itemize}[leftmargin=*]
  \checkedbox{current practices and measurements used to assess noise in the NICU}
  \checkedbox{limitations of traditional SPL measurements}
\end{itemize}
\fi

Unlike established standards for acoustic characterisation of performance spaces \citep{ISO3382-1} and open-plan offices \citep{ISO3382-3}, as well as airborne \citep{ISO16283-1}, impact \citep{ISO16283-2} and facade sound insulation \citep{iso12913-3}, there is currently a lack of internationally recognised standards for acoustic measurements in healthcare institutions. This absence of established measurement guidelines, including accuracy and placement of acoustic measurement equipment, adds to the ambiguity surrounding sound level guidelines specific to NICUs \citep{Philbin2017}. 

In the literature, NICU sound levels are commonly measured in the centre of the room \citep{Darcy2008,Lahav2015,Parra2017}, which may not accurately represent the sound levels experienced at the ears of the neonates, as the sound sources are typically situated near their head-area \cite{Darbyshire2019}. To estimate the sound levels experienced by neonates more effectively while minimising disruption to clinical operations, some studies have placed sound level meters (SLMs) \citep{Krueger2007,Romeu2016,Shoemark2016}, noise dosimeters \citep{Parra2017,Smith2018}, or microphones \citep{Park2017neonatal,Bertsch2020,Reuter2023} near the neonates' head area at varying distances. However, due to the dynamic nature of the critical care environment, the sound field surrounding the measurement instruments may inadvertently be influenced by reflective surfaces and proximity to noise sources, such as respirators and CPAP machines. Consequently, the most accurate measurement of neonates' noise exposure can be obtained by measuring sound levels very close to the opening of their ears or within the ear canal itself on actual patients \citep{Surenthiran2003} or on neonate simulators \citep{Bertsch2020,Reuter2023}. 

 Regarding measurement accuracy, traceable calibration to international standards ensures repeatability and reliability of acoustic measurements. This involves adhering to standards such as IEC~61672\nobreakdash-1 classification for SLMs and noise dosimeters \citep{IEC6167212013}, as well as the IEC~61094\nobreakdash-1 and IEC~61094\nobreakdash-4 standards for laboratory and working standard measurement microphones, respectively \citep{IEC61904-1,IEC6109441996}. Achieving similar accuracy with SLMs and dosimeters in the measurement of SPLs in a random-incidence field requires the use of measurement microphone diaphragms that are as small as possible ($\le\SI{13}{\milli\meter}$) or up to \SI{26}{\milli\meter} with random-incidence correction \citep{ISO3382-1}.

\subsection{Research questions} \label{sec:rq}

To address the gaps in determining sensible acoustic guidelines for NICUs, we investigated the microphone placements through a long-term measurement campaign in an operational neonatal unit. Specifically, the following research questions were addressed:
\begin{itemize}
    \item[RQ1] Do different microphone positions influence the measurement of noise levels in neonatal critical care environments?
    \item[RQ2] What role do bed position and ward layout play in shaping neonatal noise exposure in critical care units?
    \item[RQ3] How suitable are additional (psycho)acoustic metrics for assessing the complex sound environment in neonatal critical care?
\end{itemize}

\iftodo
\begin{itemize}[leftmargin=*]
  \checkbox{novelty and significance of measuring and analyzing sound at the ears of a neonate} 
  \checkbox{no self-noise from infant by using the simulator doll}
  \checkbox{expanded analysis on acoustic indices to analyse low-frequencies (C-weighting), Tonality indices (e.g. regression with SPL indices to understand contribution/correlation with alarms), OR(N) to understand frequency of threshold exceeded for loud events}
\end{itemize}
\fi

\section{Method}
\label{sec:Method}

\subsection{Site characteristics and administration}

\iftodo
\begin{itemize}[leftmargin=*]
    \checkedbox{Measurement site}
    \begin{itemize}
        \checkedbox{Floor plan of NICU and HD}
        \checkedbox{Capacity and features e.g. open area, no of beds, isolation wards etc} See: \citep{Mayhew2022}
        \checkedbox{Current noise mitigation measures: SoundSign}
    \end{itemize}
    \checkedbox{NICU \& HD ward occupancy}
    \begin{itemize}
        \checkedbox{Bed occupancy rate (preferably daily/weekly)}
        \checkedbox{Staff availability rate (preferably daily/weekly)}
    \end{itemize}
    \checkedbox{NICU \& HD ward schedule}
    \begin{itemize}
        \checkedbox{Visitation schedule}
        \checkedbox{Feeding schedule}
        \checkedbox{Staff shift schedule}
    \end{itemize}
\end{itemize}
\fi

Singapore General Hospital serves as the largest tertiary hospital in Singapore and houses the Department of Neonatal and Development Medicine, offering a comprehensive spectrum of services for newborns, ranging from standard to intensive care. Within this department, there is a dedicated newborn nursery, a neonatal high dependency (HD) unit, and a neonatal intensive care unit (NICU).

This study specifically focused on the neonatal HD and NICU units. The HD unit comprises two rooms housing a total of 18 cots, where neonates classified as level 2 under the Provincial Council for Maternal and Child Health Standardized Levels of Care infant acuity levels receive specialized care. Simultaneously, the NICU encompasses 10 incubators or warmers, tending to neonates categorized as level 3.

The units were equipped with a comprehensive array of medical resources, including a central monitoring station, refrigerators (1 for medication and another 2 for breast milk (fresh and frozen)), computers on wheels, ventilators, warmers, infusion pumps, oxygen-air proportioners, phototherapy lights, ultrasound machine, EEG, cooling machines, and nitric oxide delivery systems. These resources collectively facilitate extensive care services for both term and preterm newborns, with each NICU bed equipped with a bedside monitoring device. 

\begin{figure*}[h]
    \centering
    \includegraphics[width=\textwidth]{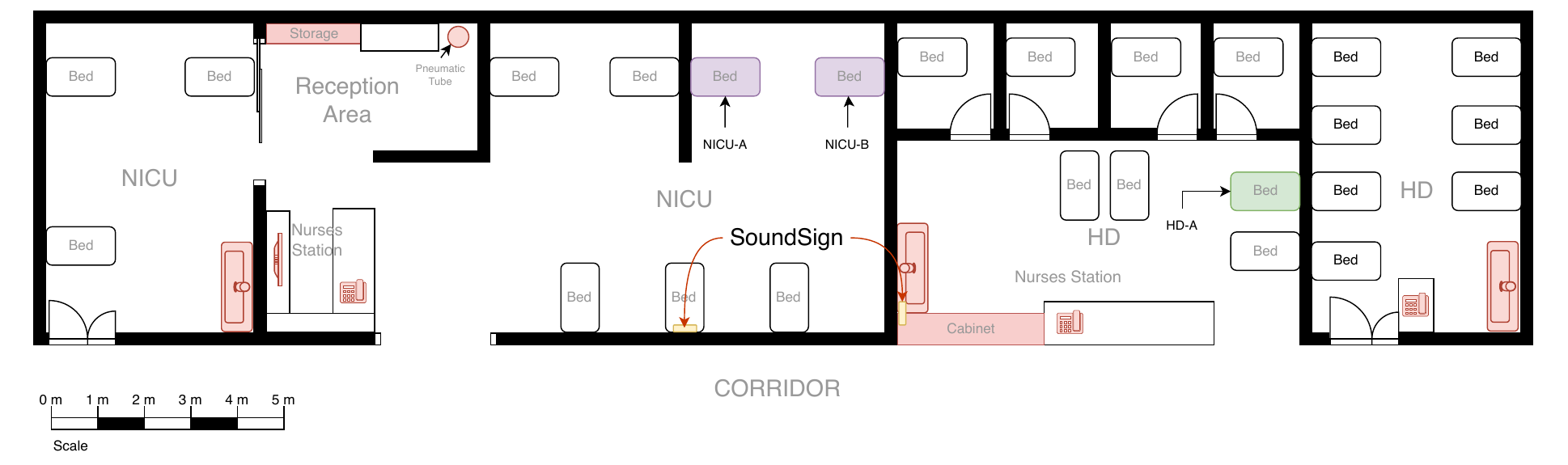}
    \caption{Floor plan depicting the layout of the neonatal intensive care unit (NICU) and high dependency (HD) ward. Measurement positions within the NICU are highlighted in violet, while the measurement point in the HD ward is marked in green. In both the NICU and HD wards, a SoundSign device was strategically positioned high on the wall.}
    \label{fig:wardlayout}
\end{figure*}

\chadded[comment=R3.2]{To prioritize patient care, the least utilized bedspaces were chosen as measurement locations. In both the NICU and HD wards, a primary bedspace was designated for measurements and labeled with the suffix ``A,'' as depicted in \Cref{fig:wardlayout}. Secondary bedspaces were labeled sequentially, e.g., NICU-B. Potential noise sources, such as wash basins, telephones, nurse stations, storage cabinets, and pneumatic tubes, were identified and marked in \Cref{fig:wardlayout}. Due to the critical condition of NICU patients, medical equipment sounds—such as ventilators and alarms—were more prevalent in the NICU compared to the HD wards. It is important to note that measurements captured only noises external to the bedspace, i.e. no medical device sounds were simulated.}

Maintaining a nurse-to-patient ratio of 1:1.5 for NICU neonates and 1:4 for HD neonates, the unit sustains a robust staffing structure with 10 nurses during day and evening shifts and 7 nurses during the night shift. Additionally, the space accommodates various healthcare professionals not exclusively stationed in the unit but utilizing the facility to assess and administer treatments to the infants.

The average patient occupancy rate for both the NICU and HD wards was approximately \SI{60}{\percent}. Notably, there were no restrictions imposed on the visitation schedule, even during the COVID-19 pandemic. However, it is important to highlight that only the parents of the patients were permitted to visit the wards. Additionally, the feeding regimen followed a schedule with intervals of three hours, each session lasting approximately one hour.

In a previous effort to reduce operational noise in the NICU, two SPL-activated warning signs named ``SoundSign'' (Cirrus Research plc, North Yorkshire, UK) was installed at high visibility areas on the wall in the NICU and HD wards, as shown in \Cref{fig:wardlayout}. The SoundSign would light up continously for \SI{30}{s} upon triggering at \Lx{AS}$=\dba{65}$, which corresponds to the \Lx{ASmax} limit in the previous editions of the \textit{Recommended Design Standards for Advanced Neonatal Care}. Currently, there is no implementation of scheduled daily "Quiet Time" \chadded[comment=3.2]{or structured programs to reduce environmental noise} in the NICU or HD wards.

\subsection{Acoustic measurement: equipment and procedure}

\iftodo
\begin{itemize}[leftmargin=*]
    \checkbox{Equipment}
    \begin{itemize}
        \highlightlist{\checkedbox{NICU incubator \& HD basinet}}
        \checkedbox{Acoustic}
    \end{itemize}
    \checkbox{Procedure}
    \begin{itemize}
        \checkedbox{Measurement period}
        \checkbox{Movement of measurement points during measurement period}
        \checkbox{Measurement positions}
        \checkedbox{Calibration}
        \checkedbox{Figure of setup}
    \end{itemize}
\end{itemize}
\fi

Measurements were carried out over a period of 20 consecutive days in March 2022, using a single incubator within the NICU ward. Similarly, within the HD ward, measurements were conducted using a single basinet for 41 consecutive days between March and April 2022.

\begin{table}[ht]
    \scriptsize
    \centering
    \caption{Bed movement schedule during the measurement period}
    \label{tab:bedmovement}
    \begin{tabularx}{\linewidth}{
    lllll%
    }
    \toprule
         Ward
         & Date
         & Time 
         & Location
         & Remarks\\
    \midrule
         NICU
         & 2022-03-03
         & 15:15 
         & NICU-A
         & Designated bedspace\\

         NICU
         & 2022-03-17
         & 11:20
         & NICU-B
         & NICU-A $\rightarrow$ NICU-B\\

         NICU
         & 2022-03-21 
         & 12:45
         & NICU-A
         & NICU-B $\rightarrow$ NICU-A\\

    \midrule
        HD
         & 2022-03-03 
         & 15:15
         & HD-A
         & Designated bedspace\\

        HD
         & 2022-04-03 
         & 09:00
         & HD-B
         & HD-A $\rightarrow$ HD-B\\

        HD
         & 2022-04-03 
         & 10:00
         & HD-A
         & HD-B $\rightarrow$ HD-A\\
    \bottomrule
    
    \end{tabularx}
\end{table}

\begin{figure}[ht]
    \centering
    \subfigure[]{
    \includegraphics[width=0.45\linewidth]{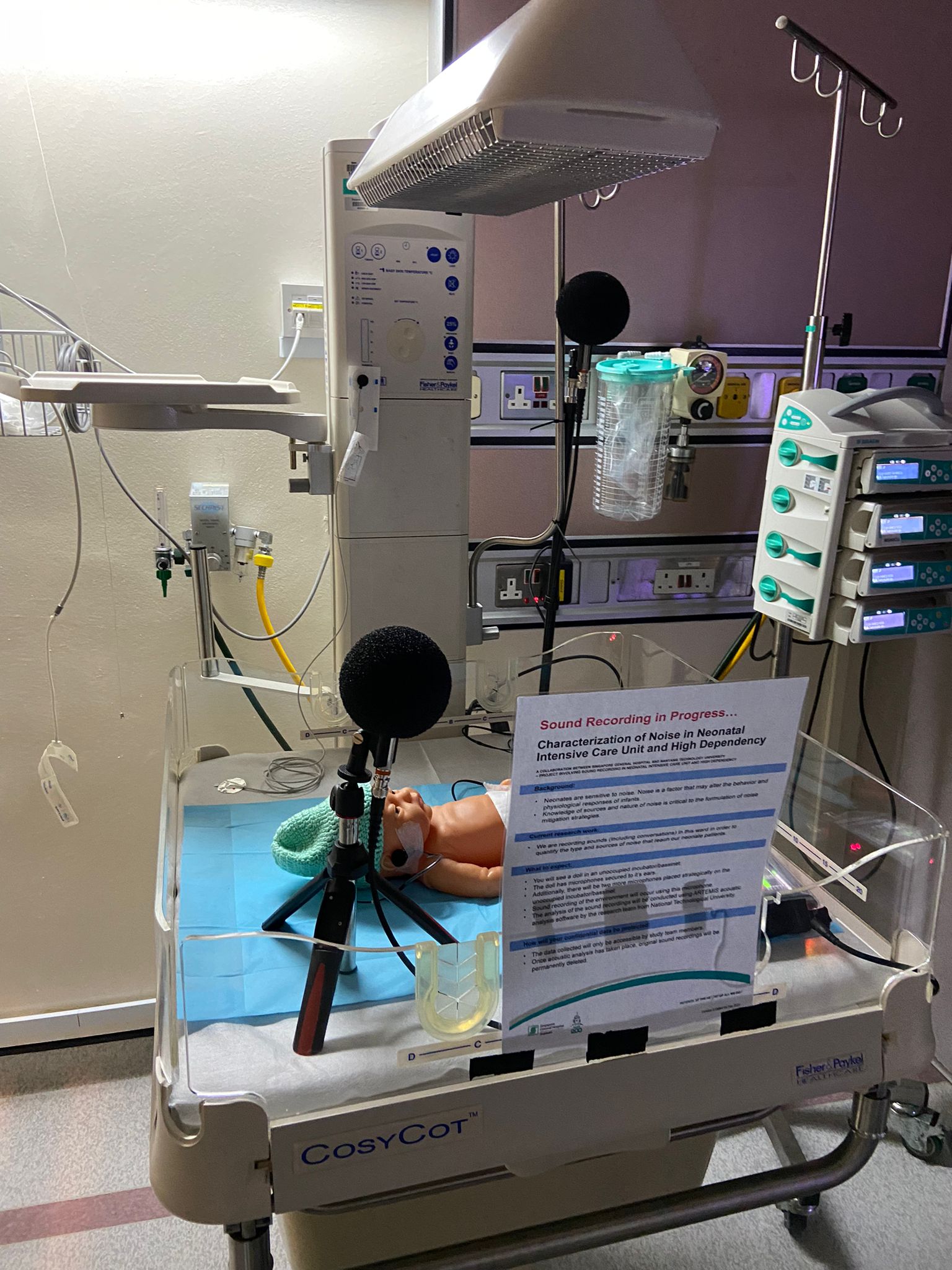}
    \label{fig:setup-incubator}}
    \subfigure[]{
    \includegraphics[width=0.45\linewidth]{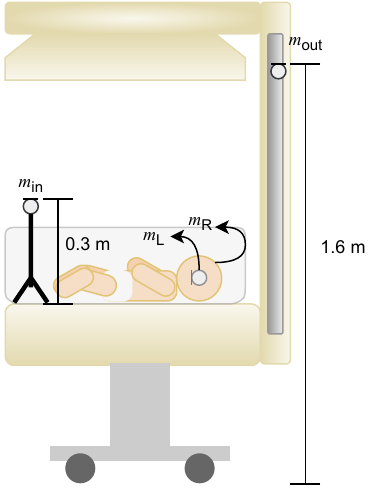}
    \label{fig:setup-incubator-mic}}
    \subfigure[]{
    \includegraphics[width=0.45\linewidth]{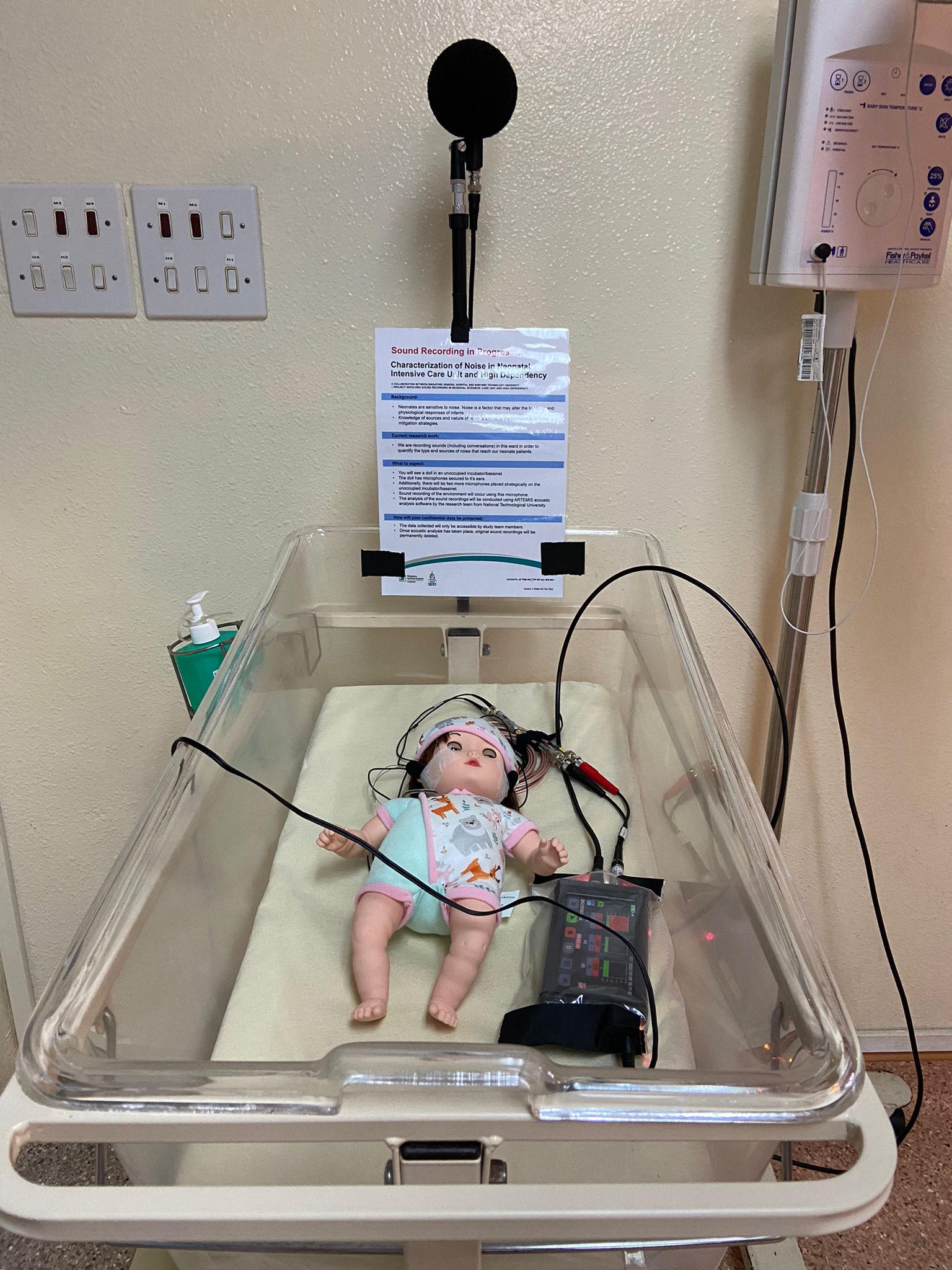}
    \label{fig:setup-basinet}}
    \subfigure[]{
    \includegraphics[width=0.45\linewidth]{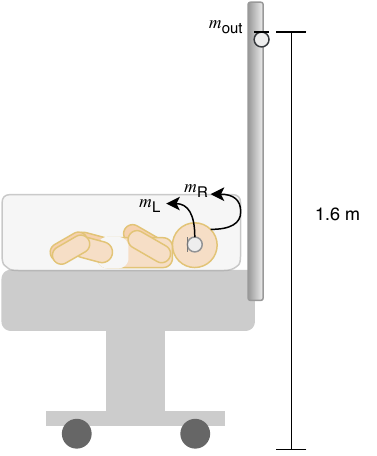}
    \label{fig:setup-basinet-mic}}
    \caption{(a) Photo and (b) diagram of the in-situ measurement setup of the incubator in the NICU, and (c) a photo and (d) diagram of bassinet setup in the high-dependency ward.}
    \label{fig:setup}
\end{figure}

To simulate how sound is perceived at the ears of neonates without capturing self-noise, calibrated binaural microphones (TYPE~4101\nobreakdash-B, Hottinger Brüel \& Kjær A/S, Virum, Denmark) were affixed on the ears of two neonate dolls with surgical tape. One doll was placed in an open-box incubator (CosyCot\textsuperscript{\textsc{tm}}, Fisher \& Paykel Healthcare Limited, Auckland, New Zealand), and the other was placed in a bassinet (Huntleigh Healthcare Limited, Wales, United Kingdom), as shown in \Cref{fig:setup}\subref{fig:setup-incubator} and \subref{fig:setup-basinet}. The left and right microphone channels of the binaural microphone would be referenced as $m_\text{L}$ and $m_\text{R}$, respectively. \chadded[comment=3.2]{Both the incubator and bassinet were open-type configurations, representative of those commonly used in the NICU and HD wards under investigation.} 

In addition, both the incubator and bassinet were fitted with a IEC~61672\nobreakdash-1 Class~1 compliant sound pressure acquisition system: IEC~61094\nobreakdash-4 WS2F microphone (146AE, GRAS Sound \& Vibration A/S, Holte, Denmark) connected to a data acquisition system (SQoBold, HEAD acoustics GmbH, Herzogenrath, Germany). To capture the sound levels around the incubator, two 146AE microphones were attached to the incubator, one secured to the IV pole \SI{1.6}{\meter} from the ground ($m_\text{out}$) and the other \SI{30}{\centi\meter} from the incubator bed, above the raised walls ($m_\text{in}$). Due to resource constraints, only one 146AE microphone was secured to the bassinet behind the head area \SI{1.6}{\meter} from the ground ($m_\text{out}$).

Before and after the measurements, each acoustic measurement device was examined to be within $\pm\SI{0.1}{\decibel}$ of \SI{94}{\decibel} and \SI{114}{\decibel} at both \SI{1}{\kilo\hertz} and \SI{250}{\hertz} using a calibrator (42AG, GRAS Sound \& Vibration A/S, Holte, Denmark). 

\subsection{(Psycho)acoustic metrics}

To evaluate the influence of microphone and bed positions on the acoustics, slow time weighted, A- and C-weighted sound pressure metrics along with their summary statistics were computed. Equivalent sound pressure levels (\Lh{AS}{1}, \Lh{CS}{1}), along with summary statistics of maximum levels (\Lh{ASmax}{1}, \Lh{CSmax}{1}), \SI{10}{\percent} exceedance levels (\Lh{AS10}{1}, \Lh{CS10}{1}), and \SI{50}{\percent} exceedance levels (\Lh{AS50}{1}, \Lh{CS50}{1}) were the primary focus. The inclusion of C-weighted metrics examines the presence of low-frequency noise, which could be heard by neonates as low as \SI{250}{\hertz} \citep{novitski_neonatal_2007}. The summary statistics, allows for a nuanced understanding of temporal variations influenced by specific microphone positions during the measurement period. Subscript reference to the 1-\si{\hour} averaging is henceforth dropped for brevity. 

The occurrence rate, \textit{OR(N)} \citep{Bliefnick2019,Ryherd2008}, which describes the percentage of time a threshold level \textit{N} was exceeded, was adopted to evaluate the possibility that the neonate would be awakened by loud sounds. A conservative threshold of \SI{5}{\decibelA} signal-to-noise ratio (SNR) above the background noise ($\textit{SNR}<5$) was employed to mitigate the chance of awakening to less than \SI{30}{\percent} \citep{kuhn_evaluating_2011}. The $\textit{OR}_\text{SNR}^h(5)$ was computed for each one hour period in a day, $h\in\{0,1,\cdots,23\}$, where the SNR was determined as \citep{Smith2018}
\begin{equation}
    \textit{SNR} = \textit{L}_\text{ASmax,1min}^h - \textit{L}_\text{AS50}^h.
\end{equation} 
Hence, $\textit{OR}_\text{SNR}^h(5)=75$ indicates that the SNR exceeded \SI{5}{\decibelA} at least once in 45 out of 60 one minute intervals in the $h$-th hour.  

Designed to quantify tonal and modulated sounds, the tonality metric, $T$, based on Sottek’s hearing model as described in the ECMA-418-2 \citep{International2020}, could indicate the prominence of alarm sounds and potentially speech. Therefore, the prominence of tonal sounds were determined by the occurrence rate where the tonality $T$ exceeded 0.4 tuHMS \citep{International2020}, within one minute periods of the $h$-th hour, i.e. $\textit{OR}^h_T(0.4)$.

\subsection{Data analysis}

The acoustic and psychoacoustic indices were computed with a commercial software package (ArtemiS \textsc{suite}, HEAD acoustics GmbH, Herzogenrath, Germany). Decibel-based metrics and their associated statistics were computed in accordance with ISO 1996-1 guidelines \citep{ISO1996-1}. Tonality was assessed using the hearing-model variant specified in ECMA-418-2 \citet{International2020}.

Due to non-normal residual distributions as indicated by the Anderson-Darling test, differences in each decibel metric across all microphones were evaluated separately for each location using a linear mixed-effects aligned ranks transformation ANOVA (LME-ART-ANOVA). Microphone type (\HDGRAS, \HDbinL, \HDbinR, \NICUGRASIn, \NICUGRASOut, \NICUbinL, \NICUbinR) was treated as a fixed effect, while the \num{1}-\si{\hour} time intervals served as a random intercept. Similarly, within- and between-ward differences for each A- and C-weighted metric were analyzed with LME-ART-ANOVA, with bed position as the fixed effect and both the \num{1}-\si{\hour} time intervals and binaural microphone positions as random effects. Differences in occurrence rates for SNR and tonality exceedances were also assessed using LME-ART-ANOVA, with bed position as the fixed effect and the \num{1}-\si{\hour} time intervals as a random effect.

All data analyses were conducted with the R programming language (R version 4.4.1) \citep{RCoreTeam2023} on a 64-bit ARM environment. The analyses were performed with these specific R packages: KS test and BH correction, with \texttt{stats} (Version 4.4.1, \citet{RCoreTeam2023}), LME-ART-ANOVA with \texttt{ARTool} (Version 0.11.1, \citet{Kay2021}), partial omega squared effect size with \texttt{effectsize} (Version 0.8.3, \citet{Ben-Shachar2020}), and acoustic data analyses with \texttt{timetk} (Version 2.9.0, \citet{dancho_timetk_2023}) and \texttt{seewave} (Version 2.2.3, \citet{sueur_seewave_2008}).

\section{Acoustic variation between microphone positions}
\label{sec:diff_mic}

This section analyses the influence of microphone positions on the assessment of noise exposure in neonatal critical care, directly addressing the first research question (RQ1). A detailed examination of disparities in A- and C-weighted decibel metrics was conducted, following the prescribed standards in \Cref{tab:sound-standards}. The computations, employing slow time-weighting and aggregation over 1-hour intervals, spanned a continuous monitoring period of 329.5 hours, ranging from 03/03/2022 at 17:00:00 to 17/03/2023 at 10:30:00 concurrently at both NICU-A and HD-A. Crucially, this time frame remained free from disruptions related to bed changes or data collection.

Differences were evaluated for each metric with the microphone positions as the repeated measures factor in the LME-ART-ANOVA and subsequent post-hoc contrast tests, as summarised in \Cref{tab:artmicHD} and \Cref{tab:artmicNICU} for HD and NICU wards, respectively. The differences in metrics between microphone positions were further examined by computing the mean differences by
\begin{equation}
    \mu^{k,x-y}_{z}=\frac{1}{N}\sum^{N}_{n=1}(L_{z,m_x^k}(n) - L_{z,m_y^k}(n)),   
\end{equation}
where $L_{z,m_x^k}(n)$ and $L_{z,m_y^k}(n)$ refer to the decibel level indices at microphone $m_x^k$ and $m_y^k$, respectively, at the $n^\text{th}$ 1-\si{\hour} period. Here, $L_z \in \{L_\text{AS}, L_\text{CS}, L_\text{ASmax}, L_\text{CSmax}, L_\text{AS10}, L_\text{CS10}, L_\text{AS50}, \\L_\text{CS50}\}$, $\{x,y\}\in\{\text{L},\text{R},\text{out},\text{in}\}$ for $k=\text{NICU}$, $\{x,y\}\in\{\text{L},\text{R},\text{out}\}$ for $k=\text{HD}$, and $N$ is the total number of 1-\si{\hour} periods. For reference, $\mu^{{\text{HD,L}-\text{out}}}_{\text{AS50}}$ refers to the mean of the differences in \Lx{AS50} between the left binaural microphone \HDbinL\ and the standard microphone \HDGRAS\ in the HD ward.

\begin{table*}[!ht]
\scriptsize
\caption{Mean and standard deviation of the differences in 1-\si{\hour} slow time-weighted metrics at the HD-A and NICU-A bed positions measured from 03/03/2022 17:00:00 to 17/03/2022 10:30:00. Difference pairs with significant ART contrasts are indicated in \textbf{bold}.}\label{tab:mean_sd_1hr}
\begin{tabularx}{1\textwidth}{
>{\RaggedRight}p{2em}
*{9}{>{\RaggedLeft}X}
}
\toprule
 & \multicolumn{3}{c}{HD-A} & \multicolumn{6}{c}{NICU-A} \\ 
\cmidrule(lr){2-4} \cmidrule(lr){5-10}

& $m_\text{L}-m$
& $m_\text{L}-m_\text{R}$ 
& $m_\text{R}-m$ 
& $m_\text{in}-m_\text{out}$
& $m_\text{L}-m_\text{in}$
& $m_\text{L}-m_\text{out}$ 
& $m_\text{L}-m_\text{R}$
& $m_\text{R}-m_\text{in}$
& $m_\text{R}-m_\text{out}$ \\

\midrule\addlinespace[2.5pt]
\Lx{AS10} 
& \textbf{-1.62 (0.97)} 
& \textbf{-0.17 (0.50)} 
& \textbf{-1.45 (0.87)} 
& \textbf{-0.47 (0.46)} 
& \textbf{2.90 (0.56) }
& \textbf{2.44 (0.54)} 
& -0.00 (0.65) 
& \textbf{2.91 (0.46)} 
& \textbf{2.44 (0.61)} \\ 

\Lx{AS50} 
& \textbf{-1.24 (0.71)} 
& \textbf{-0.20 (0.24)}
& \textbf{-1.04 (0.66)} 
& \textbf{-0.36 (0.42)} 
& \textbf{3.29 (0.58)}
& \textbf{2.93 (0.57)}
& \textbf{0.11 (0.70)}
& \textbf{3.17 (0.43)}
& \textbf{2.82 (0.56)} \\ 

\Lx{AS} 
& \textbf{-1.34 (1.18)} 
& -0.07 (0.82) 
& \textbf{-1.27 (0.79)} 
& \textbf{-0.38 (0.40)} 
& \textbf{3.02 (0.54)} 
& \textbf{2.64 (0.50)}
& 0.04 (0.55)
& \textbf{2.98 (0.38)}
& \textbf{2.60 (0.51)} \\ 

\Lx{ASmax} 
& \textbf{-1.27 (2.77)} 
& 0.04 (2.13) 
& \textbf{-1.31 (1.83)}
& \textbf{-0.59 (1.20)} 
& \textbf{3.04 (1.62)} 
& \textbf{2.44 (1.61)}
& \textbf{0.54 (1.47)} 
& \textbf{2.50 (1.39)} 
& \textbf{1.90 (1.62)} \\ 

\Lx{CS10} 
& \textbf{-0.52 (0.44)} 
& \textbf{-0.92 (0.16)}
& \textbf{0.40 (0.54)} 
& \textbf{-0.33 (0.11)} 
& \textbf{1.28 (0.35)}
& \textbf{0.95 (0.38)}
& \textbf{0.41 (0.24)}
& \textbf{0.87 (0.46)}
& \textbf{0.54 (0.50)} \\

\Lx{CS50} 
& \textbf{-0.16 (0.16)} 
& \textbf{-1.08 (0.06)} 
& \textbf{0.92 (0.19)} 
& \textbf{-0.36 (0.06)} 
& \textbf{0.94 (0.13)} 
& \textbf{0.57 (0.14)} 
& \textbf{0.54 (0.15)}
& \textbf{0.40 (0.17)} 
& 0.04 (0.19) \\ 

\Lx{CS} 
& \textbf{-0.29 (0.64)} 
& \textbf{-0.96 (0.55)} 
& \textbf{0.67 (0.36)} 
& \textbf{-0.35 (0.07)} 
& \textbf{1.09 (0.19)} 
& \textbf{0.73 (0.20)} 
& \textbf{0.48 (0.16)} 
& \textbf{0.60 (0.25)} 
& \textbf{0.25 (0.27)} \\ 

\Lx{CSmax} 
& \textbf{-1.16 (2.94)} 
& \textbf{-0.28 (1.76)} 
& \textbf{-0.88 (1.93)} 
& \textbf{-0.39 (0.79)} 
& \textbf{2.19 (0.84)} 
& \textbf{1.81 (0.98)} 
& \textbf{0.31 (0.71)} 
& \textbf{1.88 (0.87)} 
& \textbf{1.50 (1.14)} \\ 

\bottomrule
\end{tabularx}
\end{table*}

\subsection{HD ward}

For measurements at HD-A, LME-ART-ANOVA also revealed significant differences across all A-weighted decibel metrics at \SI{0.01}{\percent} significance level. Except between \binL\ and \binR\ in \Lx{AS} and \Lx{ASmax}, significant differences were found between all microphone pairs in the post-hoc contrast tests. 

A reverse trend was observed in HD-A, where the measurement microphones were louder than the binaural microphones across all A-weighted metrics between all but one binaural-measurement microphone pair. Except between \binR\ and \GRASOut\ (\mudiffLo{R}{out}{AS10}{1.45} \dba{}), mean differences across all A-weighted metrics in the rest of the binaural-measurement microphone pairs ranged from \num{-1.62} to \dba{-1.04}, as shown in \Cref{tab:mean_sd_1hr}. Hence, on average, the sound level that was exceeded \SI{10}{\percent} of the time was louder at the right ear than the measurement microphone \SI{1.6}{\meter} from the ground above the bassinet in HD-A. 

For each C-weighted decibel metric, main effects reached significance with LME-ART-ANOVA at a \SI{0.01}{\percent} significance level, indicating a large effect size. Post-hoc contrast tests further demonstrated significant differences across all microphone pairs for each C-weighted metric at HD-A, also at a \SI{0.01}{\percent} significance level.

Mean differences between \binL\ and \binR\ in C-weighted metrics were significant but small: \mudiffLo{L}{R}{CS}{-0.96}, \mudiffLo{L}{R}{CSmax}{-0.29}, \mudiffLo{L}{R}{CS10}{-0.92}, and \mudiffLo{L}{R}{CS50}{-1.08} \dbc{}. Additionally, the variability in mean differences between the binaural microphones and \GRASOut\ was more pronounced, ranging from \dLhgeC{CSmax}{1}{1.91}{6.41} and \dLhgeC{CS10}{1}{0.56}{1.56}, while remaining comparable in \dLhgeC{CS}{1}{0.46}{0.71} and \dLhgeC{CS50}{1}{0.92}{1.09}. 

Between binaural channels, low-frequency sounds were slightly louder at the right ear (\binL) across all C-weight metrics: \mudiffLo{L}{R}{CS}{-0.96}, \mudiffLo{L}{R}{CSmax}{-0.29}, \mudiffLo{L}{R}{CS10}{-0.92}, and \mudiffLo{L}{R}{CS50}{-1.08} \dbc{}. As compared to the standard measurement microphone, C-weighted metrics were slightly higher than the left but slightly lower than the right binaural channel, with the exception of the mean difference in \Lx{CSmax} levels between \binR\ and \GRASOut: \mudiffLo{R}{out}{CSmax}{-0.88} \dbc{}.

\subsection{NICU ward}

The non-parametric LME-ART-ANOVA revealed significant differences across \Lx{AS}, \Lx{ASmax}, \Lx{AS10} and \Lx{AS50} at \SI{0.01}{\percent} significance level. Subsequent pairwise post-hoc ART contrast tests indicated significant differences among all microphone pairs, except between the left and right binaural channels (\binL\ and \binR) in \Lx{AS} and \Lx{AS10} at NICU-A.  

Overall, sound levels were higher at the binaural microphones than the standard measurement microphones, indicated by mean differences across all A-weighted metrics. Mean differences in \Lx{AS} among all binaural-standard microphone pairs ranged between \mudiffLo{R}{out}{AS}{2.60} and \mudiffLo{L}{in}{AS}{3.02} \dba{}. For maximum levels, mean difference varied from \mudiffLo{R}{out}{ASmax}{1.90} to \mudiffLo{L}{in}{ASmax}{3.04} \dba{}. Likewise, mean differences in the \SI{10}{\percent} and \SI{50}{\percent} exceedance levels fell within the range of \mudiffLo{L}{out}{AS10}{2.44} to \mudiffLo{R}{in}{AS10}{2.91} \dba{}, and \mudiffLo{R}{out}{AS50}{2.82} to \mudiffLo{L}{in}{AS50}{3.29} \dba{}, respectively.

Sound levels recorded at the standard measurement microphone outside the incubator (\GRASOut) were generally significantly louder than the standard microphone inside the incubator (\GRASIn) across all A-weighted metrics,  indicating elevated external sound sources. However, mean differences across all A-weighted metrics were notably smaller than those between binaural and measurement microphones. For instance, mean differences ranged between \mudiffLo{in}{out}{AS50}{-0.36} and \mudiffLo{in}{out}{ASmax}{-0.60} \dba{}.

Although the 1-h aggregated sound levels of higher frequencies (A-weighted) were similar across both ears (\mudiffLo{L}{R}{AS}{0.039}), temporal variation between \binL\ and \binR\ is evident in the significant differences \SI{50}{\percent} exceedance levels \mudiffLo{L}{R}{AS50}{0.11} \dba{}. \chreplaced[comment=R3.3]{The significant differences between \binL\ and \binR\ in the loudest events (\mudiffLo{L}{R}{ASmax}{0.54}) contrasted with }{Interestingly, the significant differences between \binL\ and \binR\ in the loudest events (\mudiffLo{L}{R}{ASmax}{0.54}) were contrasted by} the lack of difference in loud events that occurred for about \SI{10}{\percent} of the time (\mudiffLo{L}{R}{AS10}{-0.00}), which indicate the infrequent transient nature of the loud sounds.

Statistically significant differences emerged across all C-weighted decibel metrics through LME-ART-ANOVA at a significance level of \SI{0.01}{\percent}, with a large effect size. Subsequent post-hoc contrast tests revealed significant differences among all microphone pairs for all C-weighted decibel metrics at the \SI{0.01}{\percent} significance level, with exceptions for comparisons between \binL\ and \binR, and between \GRASOut\ and \binR, both occurring at \Lx{CS50}. In these two instances at \Lx{CS50}, differences were identified at a significance level of \SI{1}{\percent} between \binL\ and \binR, while no significant differences manifested between \GRASOut\ and \binR. 

Significant differences between \binL\ and \binR\ across all C-weighted metrics indicate that left ear experienced slightly higher low-frequency sounds on average at NICU-A: \mudiffLo{L}{R}{CS}{0.48}, \mudiffLo{L}{R}{CS10}{0.41}, \mudiffLo{L}{R}{CS50}{0.54}, and \mudiffLo{L}{R}{CSmax}{0.31} \dbc{}. Similarly, low-frequency sound levels were slightly louder outside ($m_\text{out}$) than near to ($m_\text{in}$) the infant tray on average: \mudiffLo{in}{out}{CS}{-0.35} , \mudiffLo{in}{out}{CSmax}{-0.39}, \mudiffLo{in}{out}{CS10}{-0.36}, and \mudiffLo{in}{out}{CS50}{-0.33} \dbc{}. 

Differences between the binaural microphones and the standard measurement microphones for low-frequency sounds were most pronounced for the loudest sounds, while differences were small for the rest of the C-weighted metrics. Mean difference in \Lx{Cmax} levels of binaural-standard microphone pairs ranged from \mudiffLo{R}{out}{CSmax}{1.50} to \mudiffLo{L}{in}{CSmax}{2.19} \dbc{}, whereas other C-weighted metrics ranged from \mudiffLo{R}{out}{CS}{0.25} to \mudiffLo{L}{in}{CS10}{1.28} \dbc{}. Peculiarly, sound levels that were exceeded \SI{50}{\percent} of the time were similar between the right ear and the standard microphone \SI{1.6}{\meter} above the ground: \mudiffLo{R}{out}{CS50}{0.04} \dbc{}.

\section{Acoustic variation within-between wards}
\label{sec:diff_loc}

Building on the findings from \Cref{sec:diff_mic}, acoustic differences between and within wards were further analysed using only the binaural channels. The analysis focused on three bed positions: HD-A, NICU-A, and NICU-B, omitting HD-B due to the limited duration of data collection, as noted in \Cref{tab:bedmovement}. For each A- and C-weighted metric, LME-ART-ANOVA was performed with bed position as a fixed effect, and microphones and time as random effects. Post-hoc contrast tests were conducted for results with significance at the \SI{1}{\percent} level.

\subsection{A- and C-weighted metrics}

The LME-ART-ANOVA revealed statistically significant differences between bed positions for all A- and C-weighted metrics at the \SI{1}{\percent} significance level. Large effect sizes were consistently observed across most metrics, except for \Lx{ASmax} and \Lx{CSmax}, which exhibited small and medium effect sizes, respectively, as summarised in \Cref{tab:artloc}. In the post-hoc contrast tests for A-weighted metrics, significant differences were found across all bed position pairs with large effect sizes, except for HD-A vs. NICU-B in \Lx{AS10} and HD-A vs. NICU-A in \Lx{ASmax}. Similarly, post-hoc tests for C-weighted metrics indicated significant differences with large effect sizes for all bed position pairs, except for HD-A vs. NICU-B in both \Lx{CS} and \Lx{CS50}.

\begin{figure*}[h]
    \centering
    \includegraphics[width=\textwidth]{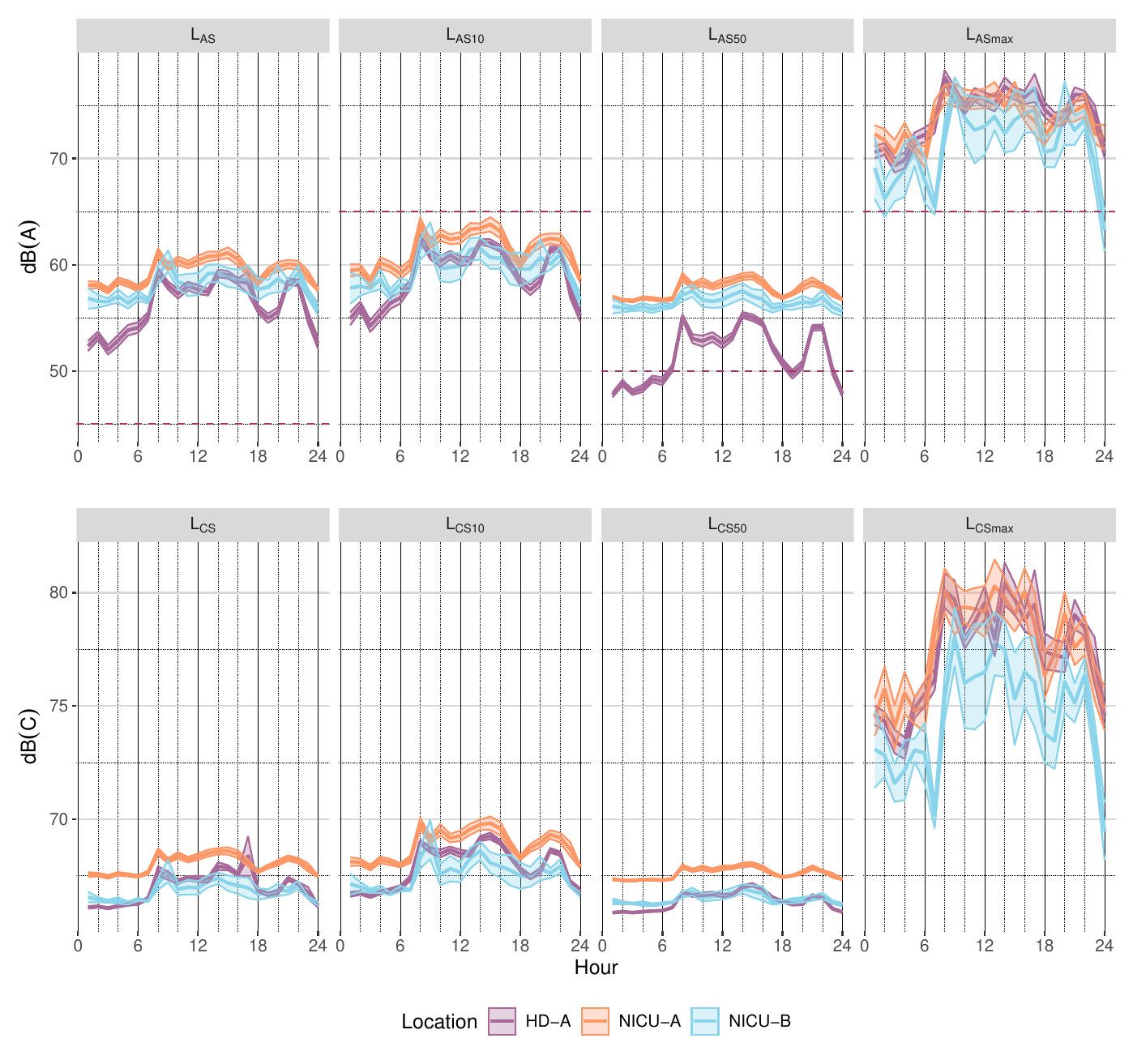}
    \caption{A- and C-weighted decibel metrics averaged by hour of the day across the entire measurement duration at NICU-A, NICU-B and HD-A measurement points.}
    \label{fig:byhourmetrics}
\end{figure*}

In general, NICU-A exhibited significantly higher A- and C-weighted metrics compared to both HD-A and NICU-B, as shown in \Cref{fig:byhourmetrics}, Between HD-A and NICU-B, \Lx{AS}, \Lx{AS50}, and \Lx{CS50} levels were higher at NICU-B, while \Lx{ASmax}, \Lx{CS10}, and \Lx{CSmax} were higher at HD-A.

\begin{figure}[h]
    \centering
    \includegraphics[width=\linewidth]{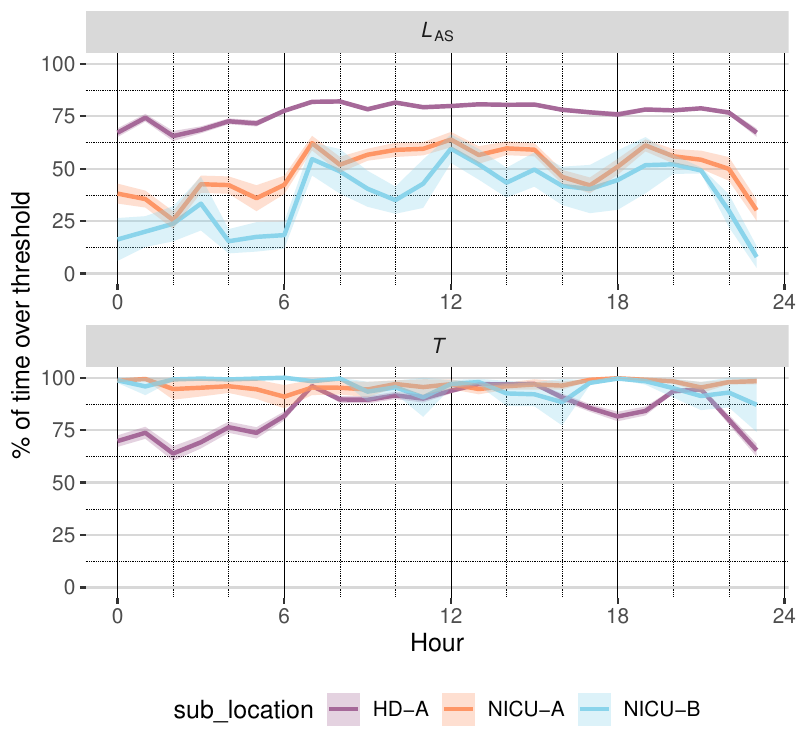}
    \caption{Occurrence rate of $\textit{OR}^h_\text{SNR}(5)$ and $\textit{OR}_{T}^h(0.4)$ averaged over the same daily 1-h period throughout the entire measurement campaign. A-weighted decibel metrics and tonality metrics averaged by hour of the day across the entire measurement duration at NICU-A, NICU-B and HD-A measurement points.}
    \label{fig:ORbyhour}
\end{figure}

\subsection{Acoustic guidelines}

Measurements of \Lx{AS10} and \Lx{AS50} across HD-A, NICU-A, and NICU-B were evaluated primarily using the 9\textsuperscript{th} and 10\textsuperscript{th} editions of the CC NICU design recommendations, as outlined in \Cref{tab:sound-standards}. For comparison, additional assessments were conducted based on \Lx{AS} and \Lx{ASmax} using the 8\textsuperscript{th} edition guidelines.

At HD-A, compliance with the \Lx{AS10} guideline was achieved for up to \SI{97}{\percent} of the $N=981$ 1-hour measurement periods, while the \Lx{AS50} limits were up to \SI{40}{\percent} compliant, as shown in \Cref{tab:reg_subloc}. Compliance rates for \Lx{AS10} and \Lx{AS50} varied depending on the microphone type, with binaural measurements showing up to \SI{8}{\percent} higher compliance for \Lx{AS10} and up to \SI{9}{\percent} higher compliance for \Lx{AS50}.

At NICU-A, compliance with the \Lx{AS10} guideline was achieved for up to \SI{99}{\percent} of the $N=492$ 1-hour measurement periods, while the \Lx{AS50} limits were consistently exceeded. Notably, microphone type influenced compliance, with binaural measurements resulting in up to \SI{8}{\percent} lower compliance for \Lx{AS10}.

At NICU-B, a similar trend to NICU-A was observed regarding adherence to \Lx{AS10} and \Lx{AS50} guidelines across the $N=98$ 1-hour measurement periods, although compliance rates remained consistent across different microphone types.

The \Lx{ASmax} and \Lx{AS} guidelines from the 8\textsuperscript{th} edition were exceeded almost consistently at across HD-A, NICU-A and NICU-B, suggesting the presence of very loud events near neonates (\Lx{ASmax} $\ge \SI{65}{\decibelA}$), though these events were infrequent, occurring less than \SI{10}{\percent} of the time. 

\chadded[]{
While \Lx{AS} exceedance patterns were consistent with previous studies irrespective of microphone placement \cite{andy_systematic_2025}, \Lx{AS} levels recorded near neonates' ears in this study aligned closely with those reported in similar NICU environments \cite{Krueger2007,Smith2018}. However, \Lx{AS} levels measured within enclosed incubators in prior studies were generally up to \dba{6} higher \cite{Romeu2016,Parra2017}, as summarized in \Cref{tab:niculevels}.}

\chadded[comment=R4.3]{Compared to \cite{Krueger2007,Parra2017}, \Lx{ASmax} levels in this study were up to \dba{23} lower, while \Lx{AS10} levels were comparable. The discrepancy in \Lx{ASmax} levels may reasonably be attributed to the absence of bedside medical devices during our measurements, though the specific cause cannot be definitively determined.}

\setlength{\LTpost}{0mm}
\begin{table*}[ht]
\scriptsize
\centering
\caption{Summary of mean A-weighted metrics and percentage of time where the metrics were within \textit{CC} guidelines over $N=981$, $N=392$, and $N=98$ 1-\si{\hour} periods at HD-A, NICU-A and NICU-B bed positions, respectively.}
\label{tab:reg_subloc}
\begin{tabularx}{\linewidth}{%
>{\raggedright\arraybackslash}p{0.04\linewidth}%
*{11}{>{\RaggedLeft}X}
}
\toprule
& \multicolumn{3}{c}{\textbf{HD-A}} 
& \multicolumn{4}{c}{\textbf{NICU-A}} 
& \multicolumn{4}{c}{\textbf{NICU-B}} \\ 

\cmidrule(lr){2-4} \cmidrule(lr){5-8} \cmidrule(lr){9-12}
& \GRASOut\textsuperscript{\textit{1}} 
& \binL\textsuperscript{\textit{1}} 
& \binR\textsuperscript{\textit{1}} 
& \GRASIn\textsuperscript{\textit{2}} 
& \GRASOut\textsuperscript{\textit{2}} 
& \binL\textsuperscript{\textit{2}} 
& \binR\textsuperscript{\textit{2}} 
& \GRASIn\textsuperscript{\textit{3}} 
& \GRASOut\textsuperscript{\textit{3}} 
& \binL\textsuperscript{\textit{3}} 
& \binR\textsuperscript{\textit{3}} \\

\midrule\addlinespace[2.5pt]
\Lx{AS} & 57.18 (3.64) & 56.05 (3.87) & 56.09 (3.48) & 56.45 (1.93) & 56.78 (2.09) & 59.33 (1.90) & 59.38 (1.90) & 55.05 (1.88) & 55.87 (1.78) & 57.47 (2.01) & 58.29 (1.81) \\ 
\Lx{AS10} & 60.25 (4.09) & 58.74 (3.85) & 58.98 (3.87) & 58.54 (2.73) & 58.96 (2.86) & 61.33 (2.65) & 61.40 (2.70) & 56.73 (2.54) & 57.55 (2.33) & 59.15 (2.74) & 59.89 (2.27) \\ 
\Lx{AS50} & 52.48 (3.71) & 51.35 (3.30) & 51.59 (3.36) & 54.53 (1.43) & 54.79 (1.55) & 57.61 (1.43) & 57.63 (1.38) & 53.47 (1.10) & 54.60 (1.27) & 55.97 (1.44) & 56.80 (1.15) \\ 
\Lx{ASmax} & 74.85 (4.14) & 73.97 (5.28) & 73.76 (4.20) & 70.96 (4.00) & 71.55 (3.99) & 73.94 (4.21) & 73.40 (4.12) & 68.44 (4.85) & 68.78 (4.92) & 70.71 (5.05) & 71.47 (4.81) \\ 
\midrule
\multicolumn{12}{l}{\Lx{AS10} $\le$ \dba{65} (\textit{CC 9\textsuperscript{th}/10\textsuperscript{th} ed})}\\ 
    Yes & 877 (89\%) & 953 (97\%) & 942 (96\%) & 387 (99\%) & 386 (98\%) & 360 (92\%) & 353 (90\%) & 98 (100\%) & 98 (100\%) & 97 (99\%) & 97 (99\%) \\ 
    No & 104 (11\%) & 28 (2.9\%) & 39 (4.0\%) & 5 (1.3\%) & 6 (1.5\%) & 32 (8.2\%) & 39 (9.9\%) & 0 (0\%) & 0 (0\%) & 1 (1.0\%) & 1 (1.0\%) \\ 
\midrule
\multicolumn{12}{l}{\Lx{AS50} $\le$ \dba{50} (\textit{CC 9\textsuperscript{th}/10\textsuperscript{th} ed})}\\
    Yes & 309 (31\%) & 389 (40\%) & 368 (38\%) & 0 (0\%) & 0 (0\%) & 0 (0\%) & 0 (0\%) & 0 (0\%) & 0 (0\%) & 0 (0\%) & 0 (0\%) \\ 
    No & 672 (69\%) & 592 (60\%) & 613 (62\%) & 392 (100\%) & 392 (100\%) & 392 (100\%) & 392 (100\%) & 98 (100\%) & 98 (100\%) & 98 (100\%) & 98 (100\%) \\ 
\midrule
\multicolumn{12}{l}{\Lx{ASmax} $\le$ \dba{65} (\textit{CC 8\textsuperscript{th} ed})}\\ 
    Yes & 8 (0.8\%) & 21 (2.1\%) & 19 (1.9\%) & 24 (6.1\%) & 14 (3.6\%) & 6 (1.5\%) & 4 (1.0\%) & 24 (24\%) & 22 (22\%) & 12 (12\%) & 10 (10\%) \\ 
    No & 973 (99\%) & 960 (98\%) & 962 (98\%) & 368 (94\%) & 378 (96\%) & 386 (98\%) & 388 (99\%) & 74 (76\%) & 76 (78\%) & 86 (88\%) & 88 (90\%) \\ 
\midrule
\multicolumn{12}{l}{\Lx{AS} $\le$ \dba{45} (\textit{CC 8\textsuperscript{th} ed})}\\
    Yes & 0 (0\%) & 0 (0\%) & 0 (0\%) & 0 (0\%) & 0 (0\%) & 0 (0\%) & 0 (0\%) & 0 (0\%) & 0 (0\%) & 0 (0\%) & 0 (0\%) \\ 
    No & 981 (100\%) & 981 (100\%) & 981 (100\%) & 392 (100\%) & 392 (100\%) & 392 (100\%) & 392 (100\%) & 98 (100\%) & 98 (100\%) & 98 (100\%) & 98 (100\%) \\ 
\bottomrule
\end{tabularx}
\begin{minipage}{\linewidth}
\textsuperscript{\textit{1}}$N=981$; \textsuperscript{\textit{2}}$N=392$; 
\textsuperscript{\textit{3}}$N=98$;
Mean (SD); n (\%)\\
\end{minipage}
\end{table*}

\subsection{Occurrence rates}

Significant differences in the frequency of loud events ($\textit{OR}_\text{SNR}^h(5)$) were found across HD-A, NICU-A, and NICU-B. Post-hoc contrasts revealed that loud events occurred significantly more frequently at HD-A than at both NICU-A and NICU-B, as depicted in \Cref{fig:ORbyhour}. Although slightly higher at NICU-A than NICU-B, the difference in $\textit{OR}_\text{SNR}^h(5)$ was still statistically significant.

The hourly variation in loud event occurrences remained relatively stable throughout the measurement period, as indicated by the standard error in \Cref{fig:ORbyhour}. At HD-A, the average occurrence remained consistent throughout the day, whereas at NICU-A and NICU-B, loud events were more frequent between 7 a.m. and 9 p.m.

Significant differences were also noted in the occurrence of tonal events ($\textit{OR}_\text{T}^h(0.4)$) across HD-A, NICU-A, and NICU-B. Post-hoc tests showed that the $\textit{OR}_\text{T}^h(0.4)$ values for NICU-A and NICU-B were statistically similar, but both were significantly higher than those for HD-A.

At NICU-A and NICU-B, tonal or modulated signals were present almost continuously ($T>0.4$), while at HD-A, such signals occurred consistently only between 7 a.m. and 9 p.m.

\section{Results and Discussion}
\label{sec:discussion}

The following discussion seeks to research questions established in \Cref{sec:rq} in the context of providing insight into the implications for clinical practice and NICU design guidelines. \Cref{sec:rq1micpos} examines how microphone positions influences the measurement of noise levels (RQ1). \chreplaced[]{
\Cref{sec:rq2bedward} explores the influence of bed positions and ward layout on neonatal auditory perception of the environment (RQ2). \Cref{sec:rq3other} reviews the suitability of non-traditional (psycho)acoustic parameters in assess the neonatal critical care acoustic environment. Finally, limitations and future work are presented in \Cref{sec:limits}
}
{
The research questions in \Cref{sec:rq} are discussed sequentially in \Cref{sec:rq1micpos}, \ref{sec:rq2bedward}, and \ref{sec:rq3other}, followed by limitations and future work in \Cref{sec:limits}.
} 

\subsection{Do different microphone positions influence the measurement of noise levels in neonatal critical care environments?} \label{sec:rq1micpos}

The results indicate that microphone positioning significantly impacts the assessment of noise exposure in neonatal critical care environments. In NICU-A, overall sound levels were consistently higher at the binaural microphones compared to the standard microphones, whereas in HD-A, the opposite trend was observed, with standard microphones capturing higher sound levels than the binaural ones. This contrast suggests that the spatial location of microphones relative to the neonate’s ears plays a crucial role in accurately capturing their auditory experience, especially in reverberant environments and dynamic noise sources.

When comparing the left and right binaural microphones, A-weighted metrics — which capture higher frequency sounds — were generally similar between both ears in both NICU-A and HD-A. However, a notable difference emerged during the loudest events (\Lx{ASmax}) in NICU-A, where the higher frequency sounds were more pronounced in one ear, indicating potential directional sound sources, such as cardiac alarms. On the other hand, C-weighted metrics, which focus on lower frequency sounds, showed consistent disparities between the left and right binaural microphones. These differences were more pronounced in HD-A than in NICU-A, suggesting that low-frequency sounds are perceived differently across both ears and are more variable in open ward environments like HD-A.

The distinctions observed in C-weighted metrics can be attributed to the significant mean differences in low-frequency sound levels, as highlighted in \Cref{tab:mean_sd_1hr}. This finding highlights the importance of placing microphones at or near ear level within open incubators, particularly in the NICU, to accurately capture fluctuations in low-frequency sounds over time. Given that neonates can discriminate sounds as low as \SI{250}{\hertz}, monitoring C-weighted metrics is crucial for a comprehensive assessment of their auditory environment.

The influence of microphone positions is further evident in the varying compliance rates in the CC guidelines. In NICU-A and NICU-B, lower compliance was noted with binaural microphones, whereas in HD-A, the standard microphones showed reduced compliance. These findings highlight the need for future guidelines to adopt standardized protocols and equipment that more accurately reflect the soundscape experienced by neonates. For instance, ISO 12913-2 \citep{iso12913-2} mandates the use of artificial head and torso simulators (HATS) with traceable calibration for soundscape assessments. However, since HATS that replicate neonatal hearing mechanisms are not yet available, standardized neonatal mannequins equipped with calibrated binaural microphones, as employed in this study, could provide a suitable alternative. 

\chadded[comment=R3.4]{Additionally, it is crucial to acknowledge that the observations in this study pertain only to external noise sources, as self-noise, such as sounds from medical equipment, caregiving activities, or crying infants, was not captured. Here, ``crying infants'' refers specifically to the sounds that would have been recorded if the microphones were affixed to actual infants rather than neonatal dolls. This distinction ensures that the findings focus on environmental noise external to the measurement setup.} 

\subsection{What role do bed position and ward layout play in
shaping neonatal noise exposure in critical care units?}
\label{sec:rq2bedward}

Measurements revealed that bed position and ward layout considerably influence neonatal noise exposure. The differences between NICU-A, NICU-B, and HD-A bed positions were statistically significant across both A- and C-weighted metrics, with large effect sizes in most cases. NICU-A consistently had the highest noise levels across metrics, likely due to its proximity and frequency of noise sources such as cardiac monitor alarms and personnel activities of the reception area in the vicinity. This suggests that NICU design and bed placement can play a significant role in either mitigating or exacerbating noise exposure for neonates.

In contrast, NICU-B exhibited lower noise levels than NICU-A across most metrics but was still louder than HD-A in specific instances, such as for \Lx{AS}, \Lx{AS50}, and \Lx{CS50}. The relative quietness in HD-A may be attributed to its spatial configuration, which is more isolated from noise sources, or differences in staff activity levels, or reduced reverberance. These findings emphasise the need for strategic bed placement and ward design to minimise harmful noise exposure, considering both the location of noise sources and the acoustic properties of the ward environment.

\subsection{How suitable are additional (psycho)acoustic metrics for assessing the complex sound environment in
neonatal critical care?} \label{sec:rq3other}

While traditional A-weighted metrics remain useful, the results suggest that incorporating additional (psycho)acoustic metrics such as tonality and signal-to-noise ratio can offer a more comprehensive evaluation of the sound environment in neonatal critical care. The differences observed in the frequency of loud events and tonal occurrences across bed positions and wards highlights the variability in how noise manifests, both temporally and spectrally.

For instance, the NICU wards exhibited continuous tonal signals ($T>0.4$), which were more frequent than in the HD ward. Such signals, though not necessarily loud, could contribute to sensory overload for neonates and affect their development if persistent \citep{lejeune_sound_2016}. Loud event occurrence rates were higher at HD-A than at both NICU-A and NICU-B, suggesting that while average sound levels may be lower, transient loud noises are more frequent, potentially disrupting sleep and physiological outcomes \citep{kuhn_evaluating_2011}. 

These findings indicate that integrating (psycho)acoustic metrics with standard decibel-based measurements could lead to a more nuanced understanding of the neonatal soundscapes. This is particularly relevant when designing interventions or making decisions on NICU layouts, as it aligns more closely with the perceptual impact of noise on neonates, potentially leading to better health outcomes.

\subsection{Limitations and future work} \label{sec:limits}

This study provides valuable insights into noise assessment in neonatal critical care units (NICUs), yet several limitations warrant consideration. Although binaural microphones offer a closer approximation of what neonates might experience, they do not fully replicate the unique auditory anatomy of neonates. The absence of standardised HATS designed specifically for neonates limits the generalisability and replicability of these findings, given that existing HATS are calibrated for adult sound perception. Developing neonatal-specific HATS or alternative simulation models could bridge this gap in future research.

Second, owing to operational constrains, the study only assessed soundscapes in three bed positions across two different wards, which may not fully capture the variability in noise exposure across diverse neonatal intensive care unit (NICU) layouts and configurations. Expanding this research \chreplaced[comment=R4.1]{to encompass varied ward layouts and larger number of measurement points is needed to enhance generalizability of these findings and}{to include more varied environments and larger sample sizes} would provide a broader understanding of how ward design, equipment placement, and caregiving activities influence noise exposure.

Lastly, while this study incorporated (psycho)acoustic metrics such as tonality and transient event occurrences, it is important to note that the thresholds used were primarily derived from adult hearing models (e.g., $T>0.4$ from \citep{International2020}) or limited empirical evidence (e.g., $\text{SNR}>\dba{5}$ for possible waking events in \citep{kuhn_evaluating_2011}). Future research should focus on establishing neonatal-specific thresholds and models for (psycho)acoustic metrics, taking into account their unique auditory profiles and developmental stages.

\section{Conclusion and recommendations}

Despite the existence of specific acoustic guidelines for neonatal intensive care units (NICU), no internationally recognised standards for acoustic measurements in healthcare settings currently exist. This has inadvertently resulted in notable disparities in how NICU soundscapes are assessed across the literature, particularly in terms of instrument type and placement, resulting in reduced reliability and comparability of the results. Through a detailed measurement campaign in an operational NICU and high dependency (HD) ward, this work investigated the impact of microphone type and placement, bed positioning, and ward layout on the assessment of neonatal soundscapes. To investigate the potential limitations of ubiquitous A-weighted decibel metrics, neonatal soundscapes were further assessed with other (psycho)acoustic parameters such as C-weighted sound pressure level, tonality and signal-to-noise ratio.

Assessment of microphone positions across all (psycho)acoustic parameters within each ward revealed significant differences between standard measurement and binaural microphones, which were affixed to the ears of a neonate doll. With differences also occurring for loud events (\Lx{ASmax}) and low-frequency sounds (C-weighted metrics) between binaural channels, it further indicates that binaural microphone placements would provide a more representative aural experience for neonates. Furthermore, significant differences between bed positions in NICU and across wards lend further support to binaural monitoring (with artificial simulators) for more holistic assessment in dynamic neonatal critical care environments. Notably, the need for additional (psycho)acoustic metric assessment is evidenced in the high occurrence rates for loud events occurring \dba{5} above the background noise levels and prominent tonal events that were not captured by A-weighted decibel metrics.

Evidence gathered in this study point towards a pressing need for further investigation and development of \chadded[comment=R3.5 R4.5]{standardized international} acoustic measurement protocols and instruments that accurately capture the aural experience of vulnerable neonates \chreplaced[comment=R3.5 R4.5]{This includes employing binaural measurements with standardized neonate dolls and incorporating additional metrics, such as C-weighted levels and tonality. These enhancements offer actionable insights for tailoring acoustic treatments and care protocols, aiming to minimize loud events, reduce low-frequency noise propagation, and improve the accessibility and audibility of maternal voice, which is essential for neonatal development.}{such as the use binaural measurement with standardised neonate dolls}.

\chadded[]{Implementation of the proposed measurement techniques and guidelines would require close collaboration between engineers and clinicians during initial design and construct of the NICU environment, in order to optimize positioning of cots and equipment, balancing between acoustic quality, infection prevention measures and work efficiency. Collaboration between medical device companies, hospital engineers and clinicians is also crucial in designing future devices or modules suitable for the NICU patient population.} 

\chadded[comment=R4.6]{In addition is a need for management commitment to invest in monitoring equipment and regular conduct of such exercises to allow for continuous improvement in a timely manner.}

\section*{Data Availability}
The data that support the findings of this study are openly available in NTU research data repository DR-NTU (Data) at \url{https://doi.org/10.21979/N9/8GHNGX}, and replication code used in this study is available on GitHub at the following repository: \href{https://doi.org/10.5281/zenodo.14643228}{https://doi.org/10.5281/zenodo.14643228}.

\section*{Declaration of competing interest}
The authors declare that they have no known competing financial interests or personal relationships that could have appeared to influence the work reported in this paper.

\section*{Acknowledgments}
This work was made possible through a research collaboration agreement between Nanyang Technological University and Singapore General Hospital.

\printcredits

\bibliographystyle{model1-num-names}
\bibliography{references_zot}

\setcounter{section}{0}
\renewcommand{\thesection}{Appendix \Alph{section}}

\setcounter{table}{0}
\renewcommand{\thetable}{\Alph{section}.\arabic{table}}

\setcounter{figure}{0}
\renewcommand{\thefigure}{\Alph{section}.\arabic{figure}}
%\appendix

\onecolumn

\clearpage

\section{Statistical test results} \label{sec:stats}

\setlength{\LTpost}{0mm}
\setcounter{table}{0}

%\begin{table}[!b]
%\footnotesize
\scriptsize

\begin{xltabular}{0.9\textwidth}{
>{\raggedright\arraybackslash}p{0.1\textwidth}
>{\raggedright\arraybackslash}p{0.1\textwidth}
>{\raggedright\arraybackslash}p{0.25\textwidth}
>{\raggedleft\arraybackslash}p{0.4\textwidth}
>{\raggedleft\arraybackslash}p{0.05\textwidth}
}
\caption{Summary of \lmeart\ and posthoc contrast tests with microphone type as the fixed effect, and \num{1}-\si{\hour} time periods as the random effect for each acoustic metric at the HD ward.} \label{tab:artmicHD} 
\\
\toprule
\textbf{Metric}
& \textbf{Term} 
& \textbf{Test}\textsuperscript{\textit{1}} 
& $p$-\textbf{value}\textsuperscript{\textit{2}} 
& $\omega^2_\text{P}$\textsuperscript{\textit{3}} 
\\ 
\endfirsthead

\midrule\addlinespace[2.5pt]

\Lx{AS} & microphone & LME-ART-ANOVA & ****0.0000 & (L)0.57 \\ 
&   \HDGRAS -- \HDbinL & ART Contrasts & ****0.0000 &  \\ \addlinespace[2.5pt] 
&   \HDGRAS -- \HDbinR & ART Contrasts & ****0.0000 &  \\ \addlinespace[2.5pt] 
&   \HDbinL -- \HDbinR & ART Contrasts &  0.1147 &  \\ 
\midrule\addlinespace[2.5pt]
\addlinespace[2.5pt]
\Lx{ASmax} & microphone & LME-ART-ANOVA & ****0.0000 & (L)0.25 \\ 
&   \HDGRAS -- \HDbinL & ART Contrasts & ****0.0000 &  \\ \addlinespace[2.5pt] 
&   \HDGRAS -- \HDbinR & ART Contrasts & ****0.0000 &  \\ \addlinespace[2.5pt] 
&   \HDbinL -- \HDbinR & ART Contrasts &  0.7208 &  \\ 
\midrule\addlinespace[2.5pt]
\Lx{AS10} & microphone & LME-ART-ANOVA & ****0.0000 & (L)0.67 \\ 
&   \HDGRAS -- \HDbinL & ART Contrasts & ****0.0000 &  \\ \addlinespace[2.5pt] 
&   \HDGRAS -- \HDbinR & ART Contrasts & ****0.0000 &  \\ \addlinespace[2.5pt] 
&   \HDbinL -- \HDbinR & ART Contrasts & ***0.0002 &  \\ 
\midrule\addlinespace[2.5pt]
\Lx{AS50} & microphone & LME-ART-ANOVA & ****0.0000 & (L)0.74 \\ 
&   \HDGRAS -- \HDbinL & ART Contrasts & ****0.0000 &  \\ \addlinespace[2.5pt] 
&   \HDGRAS -- \HDbinR & ART Contrasts & ****0.0000 &  \\ \addlinespace[2.5pt] 
&   \HDbinL -- \HDbinR & ART Contrasts & ****0.0000 &  \\ 

\bottomrule
\end{xltabular}

%footnote
\noindent
\begin{minipage}{1\textwidth}
\textsuperscript{\textit{1}}Linear mixed effects Aligned Rank Transform (ART) ANOVA (\lmeart); \\
\textsuperscript{\textit{2}}$\text{*}p<0.05$; $\text{**}p<0.01$; $\text{***}p<0.001$; $\text{****}p<0.0001$\\
\textsuperscript{\textit{3}}Partial Omega squared ($\omega^2_\text{P}$) for linear mixed effects. (L) large effect $\omega^2_\text{P}\ge0.14$ ; (M) medium effect $0.06\ge\omega^2_\text{P}<0.14$; (S) small effect $0.01\ge\omega^2_\text{P}<0.06$ \\
\end{minipage}

%\end{table}

\setlength{\LTpost}{0mm}
\setcounter{table}{1}

%\begin{table}[!ht]
\scriptsize
\begin{xltabular}{0.9\textwidth}{
>{\raggedright\arraybackslash}p{0.1\textwidth}
>{\raggedright\arraybackslash}p{0.1\textwidth}
>{\raggedright\arraybackslash}p{0.25\textwidth}
>{\raggedleft\arraybackslash}p{0.4\textwidth}
>{\raggedleft\arraybackslash}p{0.05\textwidth}
}

\caption{Summary of \lmeart\ and posthoc contrast tests with microphone type as the fixed effect, and \num{1}-\si{\hour} time periods as the random effect for each acoustic metric at the NICU ward.} \label{tab:artmicNICU} 
\\
\toprule
\textbf{Metric}
& \textbf{Term} 
& \textbf{Test}\textsuperscript{\textit{1}} 
& $p$-\textbf{value}\textsuperscript{\textit{2}} 
& $\omega^2_\text{P}$\textsuperscript{\textit{3}}  \\ 
\endfirsthead

\multicolumn{5}{r@{}}{\small Continued from the previous page\ldots}\\

\toprule
\textbf{Metric}
& \textbf{Term} 
& \textbf{Test}\textsuperscript{\textit{1}} 
& $p$-\textbf{value}\textsuperscript{\textit{2}} 
& $\omega^2_\text{P}$\textsuperscript{\textit{3}}  \\ 
\midrule\addlinespace[2.5pt]
\endhead

\multicolumn{5}{r@{}}{\small Continues to the next page\ldots}\\
\endfoot
%blank for last footer
\endlastfoot

\midrule\addlinespace[2.5pt]
\Lx{AS} & microphone & LME-ART-ANOVA & ****0.0000 & (L)0.91 \\ \addlinespace[2.5pt] 
&   \NICUGRASIn -- \NICUGRASOut & ART Contrasts & ****0.0000 &  \\ \addlinespace[2.5pt] 
&   \NICUGRASIn -- \NICUbinL & ART Contrasts & ****0.0000 &  \\ \addlinespace[2.5pt] 
&   \NICUGRASIn -- \NICUbinR & ART Contrasts & ****0.0000 &  \\ \addlinespace[2.5pt] 
&   \NICUGRASOut -- \NICUbinL & ART Contrasts & ****0.0000 &  \\ \addlinespace[2.5pt] 
&   \NICUGRASOut -- \NICUbinR & ART Contrasts & ****0.0000 &  \\ \addlinespace[2.5pt] 
&   \NICUbinL -- \NICUbinR & ART Contrasts &  0.3087 &  \\ 
\midrule\addlinespace[2.5pt]
\Lx{ASmax} & microphone & LME-ART-ANOVA & ****0.0000 & (L)0.64 \\ \addlinespace[2.5pt] 
&   \NICUGRASIn -- \NICUGRASOut & ART Contrasts & ****0.0000 &  \\ \addlinespace[2.5pt] 
&   \NICUGRASIn -- \NICUbinL & ART Contrasts & ****0.0000 &  \\ \addlinespace[2.5pt] 
&   \NICUGRASIn -- \NICUbinR & ART Contrasts & ****0.0000 &  \\ \addlinespace[2.5pt] 
&   \NICUGRASOut -- \NICUbinL & ART Contrasts & ****0.0000 &  \\ \addlinespace[2.5pt] 
&   \NICUGRASOut -- \NICUbinR & ART Contrasts & ****0.0000 &  \\ \addlinespace[2.5pt] 
&   \NICUbinL -- \NICUbinR & ART Contrasts & ****0.0000 &  \\ 
\midrule\addlinespace[2.5pt]
\Lx{AS10} & microphone & LME-ART-ANOVA & ****0.0000 & (L)0.89 \\ \addlinespace[2.5pt] 
&   \NICUGRASIn -- \NICUGRASOut & ART Contrasts & ****0.0000 &  \\ \addlinespace[2.5pt] 
&   \NICUGRASIn -- \NICUbinL & ART Contrasts & ****0.0000 &  \\ \addlinespace[2.5pt] 
&   \NICUGRASIn -- \NICUbinR & ART Contrasts & ****0.0000 &  \\ \addlinespace[2.5pt] 
&   \NICUGRASOut -- \NICUbinL & ART Contrasts & ****0.0000 &  \\ \addlinespace[2.5pt] 
&   \NICUGRASOut -- \NICUbinR & ART Contrasts & ****0.0000 &  \\ \addlinespace[2.5pt] 
&   \NICUbinL -- \NICUbinR & ART Contrasts &  0.8877 &  \\ 
\midrule\addlinespace[2.5pt]
\Lx{AS50} & microphone & LME-ART-ANOVA & ****0.0000 & (L)0.91 \\ \addlinespace[2.5pt] 
&   \NICUGRASIn -- \NICUGRASOut & ART Contrasts & ****0.0000 &  \\ \addlinespace[2.5pt] 
&   \NICUGRASIn -- \NICUbinL & ART Contrasts & ****0.0000 &  \\ \addlinespace[2.5pt] 
&   \NICUGRASIn -- \NICUbinR & ART Contrasts & ****0.0000 &  \\ \addlinespace[2.5pt] 
&   \NICUGRASOut -- \NICUbinL & ART Contrasts & ****0.0000 &  \\ \addlinespace[2.5pt] 
&   \NICUGRASOut -- \NICUbinR & ART Contrasts & ****0.0000 &  \\ \addlinespace[2.5pt] 
&   \NICUbinL -- \NICUbinR & ART Contrasts & **0.0022 &  \\ 

\bottomrule
\end{xltabular}
%\end{table}

\setlength{\LTpost}{0mm}
\setcounter{table}{2}
\footnotesize

\begin{xltabular}{1\textwidth}{
>{\raggedright\arraybackslash}p{0.05\textwidth}
>{\raggedright\arraybackslash}p{0.2\textwidth}
>{\raggedright\arraybackslash}p{0.2\textwidth}
>{\raggedleft\arraybackslash}X
>{\raggedleft\arraybackslash}p{0.1\textwidth}
}
\caption{Summary of \lmeart\ and posthoc contrast tests with bed position as the fixed effect, and \num{1}-\si{\hour} time periods and microphone type as the random effects for each acoustic metric.} \label{tab:artloc} 
\\
\toprule
\textbf{Metric} 
& \textbf{Test} 
& \textbf{Term}
& $p$-\textbf{value} 
& $\omega^2_\text{P}$\\ 
\endfirsthead

\midrule\addlinespace[2.5pt]
\Lx{AS} & LME-ART-ANOVA & Bed position & ****0.0000 & (L)0.47 \\ 
& ART Contrasts & (HD-A) - (NICU-A) & ****0.0000 & (L)-2.16 \\ 
& ART Contrasts & (HD-A) - (NICU-B) & ****0.0000 & (L)-0.92 \\ 
& ART Contrasts & (NICU-A) - (NICU-B) & ****0.0000 & (L)1.24 \\ 
\midrule
\Lx{AS10} & LME-ART-ANOVA & Bed position & ****0.0000 & (L)0.25 \\ 
& ART Contrasts & (HD-A) - (NICU-A) & ****0.0000 & (L)-1.34 \\ 
& ART Contrasts & (HD-A) - (NICU-B) &  1.0000 & (S)0.02 \\ 
& ART Contrasts & (NICU-A) - (NICU-B) & ****0.0000 & (L)1.36 \\ 
\midrule
\Lx{AS50} & LME-ART-ANOVA & Bed position & ****0.0000 & (L)0.81 \\ 
& ART Contrasts & (HD-A) - (NICU-A) & ****0.0000 & (L)-4.50 \\ 
& ART Contrasts & (HD-A) - (NICU-B) & ****0.0000 & (L)-3.27 \\ 
& ART Contrasts & (NICU-A) - (NICU-B) & ****0.0000 & (L)1.23 \\ 
\midrule
\Lx{ASmax} & LME-ART-ANOVA & Bed position & ****0.0000 & (S)0.04 \\ 
& ART Contrasts & (HD-A) - (NICU-A) &  1.0000 & (S)0.03 \\ 
& ART Contrasts & (HD-A) - (NICU-B) & ****0.0000 & (L)0.91 \\ 
& ART Contrasts & (NICU-A) - (NICU-B) & ****0.0000 & (L)0.88 \\ 
\midrule
\Lx{CS} & LME-ART-ANOVA & Bed position & ****0.0000 & (L)0.48 \\ 
& ART Contrasts & (HD-A) - (NICU-A) & ****0.0000 & (L)-2.27 \\ 
& ART Contrasts & (HD-A) - (NICU-B) &  0.5197 & (M)0.13 \\ 
& ART Contrasts & (NICU-A) - (NICU-B) & ****0.0000 & (L)2.40 \\ 
\midrule
\Lx{CS10} & LME-ART-ANOVA & Bed position & ****0.0000 & (L)0.34 \\ 
& ART Contrasts & (HD-A) - (NICU-A) & ****0.0000 & (L)-1.64 \\ 
& ART Contrasts & (HD-A) - (NICU-B) & ****0.0000 & (L)0.56 \\ 
& ART Contrasts & (NICU-A) - (NICU-B) & ****0.0000 & (L)2.20 \\ 
\midrule
\Lx{CS50} & LME-ART-ANOVA & Bed position & ****0.0000 & (L)0.54 \\ 
& ART Contrasts & (HD-A) - (NICU-A) & ****0.0000 & (L)-2.62 \\ 
& ART Contrasts & (HD-A) - (NICU-B) & *0.0290 & (L)-0.23 \\ 
& ART Contrasts & (NICU-A) - (NICU-B) & ****0.0000 & (L)2.40 \\ 
\midrule
\Lx{CSmax} & LME-ART-ANOVA & Bed position & ****0.0000 & (M)0.06 \\ 
& ART Contrasts & (HD-A) - (NICU-A) & ****0.0000 & (L)-0.28 \\ 
& ART Contrasts & (HD-A) - (NICU-B) & ****0.0000 & (L)1.05 \\ 
& ART Contrasts & (NICU-A) - (NICU-B) & ****0.0000 & (L)1.33 \\ 
\bottomrule
\end{xltabular}





\end{document}